\documentclass[aps,prx,reprint,superscriptaddress]{revtex4-2}

\usepackage[utf8]{inputenc}
\usepackage[english]{babel}
\usepackage[T1]{fontenc}
\usepackage{color}
\usepackage{xcolor}
\usepackage{microtype}
\usepackage{enumitem}
\usepackage[normalem]{ulem}
\usepackage[pdftex]{graphicx}
\usepackage{hyperref}
\usepackage{amsmath}
\usepackage{amssymb}
\usepackage{mathtools}
\usepackage{mathrsfs}
\usepackage{stmaryrd}
\usepackage{multirow}
\usepackage{bbm}
\usepackage{bbold}
\usepackage{braket}
\usepackage{siunitx}
\usepackage[caption=false]{subfig}
\usepackage{float}
\usepackage{placeins}
\usepackage{makecell}
\usepackage[capitalize]{cleveref}
\usepackage{orcidlink}
\usepackage{tikz}

\hypersetup{%
    hypertexnames=false,
    colorlinks=true,
    allcolors=blue,
    unicode=true
}

\graphicspath{{figures/}}

\newcommand{\phantomsubfloat}[1]{%
    {%
        \captionsetup[subfloat]{farskip=0pt,captionskip=0pt}
        \captionsetup[subfigure]{labelformat=empty}
        \subfloat{#1}
    }
}

\crefname{equation}{Eq.}{Eqs.}
\Crefname{equation}{Equation}{Equations}

\crefname{figure}{Fig.}{Figs.}
\Crefname{figure}{Figure}{Figures}

\crefname{section}{Sec.}{Secs.}
\Crefname{section}{Section}{Sections}

\crefname{appendix}{Appendix}{Appendices}
\Crefname{appendix}{Appendix}{Appendices}

\makeatletter
\AddToHook{cmd/appendix/before}{\def\cref@section@alias{appendix}\def\cref@subsection@alias{appendix}}
\makeatother

\usetikzlibrary{quantikz2}

\definecolor{pauliblue}{HTML}{0076BA}
\definecolor{paulired}{HTML}{C82506}
\tikzset{sensitivity band/.style={draw=pauliblue,line width=1.4pt,line cap=butt,line join=round}}

\makeatletter
\newcommand{\zband}[2]{%
    \pgfutil@g@addto@macro\tikzcd@atendlabels{%
        \begin{pgfonlayer}{background}
            \draw[sensitivity band]
                ([yshift=-0.6mm]\tikzcdmatrixname-1-#1.center) --
                ([yshift=-0.6mm]\tikzcdmatrixname-1-#2.center);
        \end{pgfonlayer}%
    }%
}
\makeatother

\newcommand{\cx}{\mathop{\mathit{CX}}}
\newcommand{\abs}[1]{\left\lvert #1 \right\rvert}
\newcommand{\norm}[1]{\left\lVert #1 \right\rVert}

\begin{document}

\title{Bounded decoding complexity in transversal circuits with magic state injection}

\author{Lukas M. Sieberer\orcidlink{0000-0002-0163-7850}}
\email{lukas.sieberer@planqc.eu}
\affiliation{PlanQC Austria GmbH, Pembaurstra\ss e 14, 6020 Innsbruck, Austria}

\author{Giacomo Giudice\orcidlink{0000-0001-7141-2553}}
\email{giacomo.giudice@planqc.eu}
\affiliation{PlanQC GmbH, M\"{u}nchener Stra\ss e 34, 85748 Garching, Germany}

\date{\today}

\begin{abstract}
    We characterize the complexity of space-time correlated decoding for fault-tolerant quantum computing architectures that implement Clifford gates transversally and realize non-Clifford gates via magic state injection.
    In this approach, decoding graphs for logical measurements are constructed by backpropagating measurement operators through the circuit, which causes them to grow with circuit volume.
    However, the decoding volume can be substantially reduced by decoding measurements sequentially, committing corrections, and excluding syndrome data that have already been processed for earlier measurements.
    For random circuits, we find that the resulting decoding volume associated with mid-circuit measurements approaches a stationary average that is determined by the density of magic state injections.
    Large decoding instances are exponentially suppressed.
    We further find that the decoding complexity is remarkably insensitive to different choices of measurement products to be decoded, and that the same qualitative behavior persists across different circuit connectivities and Clifford gate ensembles.
    These results suggest that scalable space-time correlated decoding of large-scale fault-tolerant quantum algorithms is achievable since the magic states that make transversal architectures universal also keep them decodable.
\end{abstract}

\maketitle

\section{Introduction}

Quantum computation is widely believed to offer asymptotic speedups over classical computation for certain problems, such as quantum simulation and integer factorization~\cite{nielsen2002}.
However, the noise in real devices poses a significant barrier to solving large-scale problems, and practical implementations will almost certainly require error correction~\cite{TerhalQuantumErrorCorrection2015, CampbellRoadsFaulttolerantUniversal2017, GottesmanOpportunitiesChallengesFaultTolerant2022, EisertMindGapsFraught2025}.

Over the past decade, there has been substantial progress in the experimental realization of quantum error correction across a variety of hardware platforms, including superconducting qubits~\cite{KrinnerRealizingRepeatedQuantum2022, MarquesLogicalqubitOperationsErrordetecting2022, SivakRealtimeQuantumError2023, GoogleQuantumAISuppressingQuantumErrors2023, SundaresanDemonstratingMultiroundSubsystem2023, GuptaEncodingMagicState2024, PuttermanHardwareefficientQuantumError2025, GoogleQuantumAIandCollaboratorsQuantumErrorCorrection2025, LinSurfaceCodeLogical2026, WangSuperconductingSurfacecodeProcessor2026}, trapped ions~\cite{EganFaulttolerantControlErrorcorrected2021, Ryan-AndersonRealizationRealTimeFaultTolerant2021, PostlerDemonstrationFaulttolerantUniversal2022, SteinerGenuineMultipartiteEntanglement2026}, and neutral atom arrays~\cite{BluvsteinLogicalQuantumProcessor2024, BluvsteinFaulttolerantNeutralatomArchitecture2026, ReichardtFaulttolerantQuantumComputation2025, SalesRodriguezExperimentalDemonstrationLogical2025, ChungFaulttolerantOperationMaterials2025, ComputingQuantumErrorCorrection2026}.
These experiments have demonstrated several important milestones in fault-tolerant quantum computation, including encoding of information in logical qubits, suppression of logical error rates with increasing code distance, as well as logical operations.
In parallel, significant effort has been devoted to reducing the overhead associated with fault-tolerant architectures, both in terms of space (physical qubit count) as well as time (physical circuit depth).
In particular, tailoring error correction schemes to the native connectivity and control capabilities of specific platforms has proven to be a powerful approach to minimizing overhead.

A key development in this context is the availability of reconfigurable, effectively nonlocal connectivity on certain platforms, particularly trapped-ion~\cite{Ransford98qubitTrappedionQuantum2026} and neutral-atom processors~\cite{BluvsteinLogicalQuantumProcessor2024, EveredHighfidelityEntanglingGates2026}.
This allows physical qubits that are distant in a static code layout to interact directly, enabling transversal gates between distant logical code blocks and transversal or fold-transversal gates within a block~\cite{BombinTopologicalQuantumDistillation2006, MoussaTransversalCliffordGates2016, BreuckmannFoldTransversalCliffordGates2024}, and thereby circumventing the geometric constraints of strictly local two-dimensional architectures.
For two-dimensional topological stabilizer codes with local checks, locality-preserving logical gates are restricted to the Clifford group~\cite{BravyiClassificationTopologicallyProtected2013}, which is not sufficient for universal quantum computation~\cite{EastinRestrictionsTransversalEncoded2009}.
Nevertheless, these capabilities have been leveraged for realizing low-overhead logical processing architectures based on transversal operations~\cite{ZhouLowOverheadTransversalFault2025}, as recently highlighted in several experimental demonstrations with neutral atom and ion platforms~\cite{PostlerDemonstrationFaulttolerantUniversal2022, BluvsteinLogicalQuantumProcessor2024, Ryan-AndersonHighfidelityTeleportationLogical2024, ReichardtFaulttolerantQuantumComputation2025, BluvsteinFaulttolerantNeutralatomArchitecture2026, EveredHighfidelityEntanglingGates2026, DasuComputingManyEncoded2026}.

\begin{figure}[t]
    \includegraphics[width=\linewidth,scale=0.6]{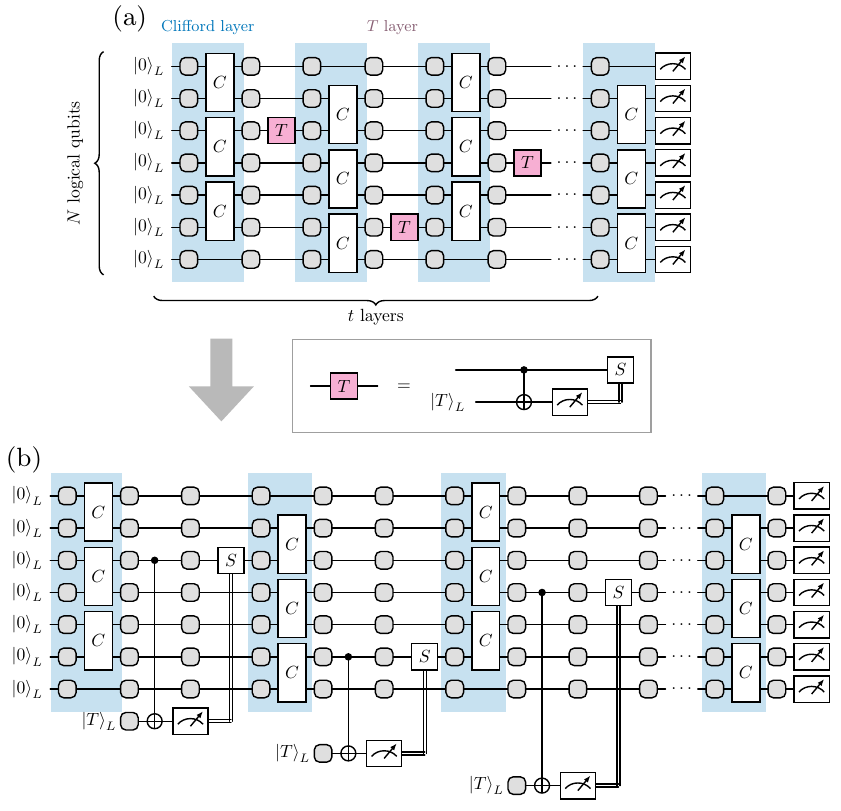}
    \vspace*{-1em}
    \caption{%
        Random-circuit model.
        (a)~$N$ computational logical qubits, initialized in $\ket{0}_L$ or $\ket{+}_L$, evolve under alternating Clifford and $T$-gate layers and are finally measured in the $X$ or $Z$ basis.
        The Clifford layers consist of random two-qubit gates in a one-dimensional brickwork geometry, while each $T$ gate is applied independently with probability $\kappa$.
        (b)~$T$ gates are implemented by gate teleportation using a resource state $\ket{T}_L$, a mid-circuit measurement, and a conditional $S$ gate.
        Grey boxes denote syndrome-extraction rounds.
    }
    \label{fig:circuit-model}
\end{figure}

An important insight emerging from these architectures is that fault tolerance can be preserved with a minimal number of syndrome-extraction rounds between logical operations.
In particular, it has been shown that a single syndrome-extraction round per transversal entangling gate, or even one round shared across several successive gates, can suffice to maintain fault tolerance under circuit-level noise models~\cite{ZhouLowOverheadTransversalFault2025, CainCorrelatedDecodingLogical2024, CainFastCorrelatedDecoding2025, WanIterativeTransversalCNOT2025, Serra-PeraltaDecodingTransversalClifford2026}.
Such a low rate of syndrome extraction rounds contrasts with the conventional prescription of performing $O(d)$ rounds to reliably distinguish data errors from measurement errors in a distance-$d$ code~\cite{DennisTopologicalQuantumMemory2002, HorsmanSurfaceCodeQuantum2012}.
The reduction in syndrome-extraction rate is enabled by \emph{space-time correlated decoding}, in which syndrome information is not decoded independently between successive logical operations but instead processed jointly across multiple circuit layers~\cite{BluvsteinLogicalQuantumProcessor2024, WanIterativeTransversalCNOT2025, ZhouLowOverheadTransversalFault2025, CainCorrelatedDecodingLogical2024, CainFastCorrelatedDecoding2025, SahayErrorCorrectionTransversal2025, ZhouLearningDecodeLogical2025, ZhangScalableConstantTimeLogical2025, Serra-PeraltaDecodingTransversalClifford2026, TurnerScalableDecodingProtocols2026, BonillaAtaidesNeuralDecodersUniversal2026, WebberHighAccuracyPropagationAssistedLocal2026}.

In Clifford circuits, correlating syndrome information across circuit layers is possible because Clifford operations map Pauli operators to Pauli operators, so that the propagation of Pauli errors and stabilizers through the logical circuit can be efficiently tracked classically via the Gottesman–Knill theorem~\cite{GottesmanHeisenbergRepresentationQuantum1998, AaronsonImprovedSimulationStabilizer2004}.
Importantly, universal quantum computation can be achieved by supplementing Clifford operations with magic state injection~\cite{BravyiUniversalQuantumComputation2005}.
The required feedforward operations are themselves Clifford, so that the non-Clifford resource is confined to the injected magic states and space-time correlated decoding can be extended to universal circuits.

By substantially reducing the number of syndrome extraction rounds, this paradigm of space-time correlated decoding directly lowers the time overhead associated with the implementation of error correction.
At the same time, the size of the decoding graph and the complexity of the associated inference problem can grow substantially when adopting a correlated decoding strategy.
This is because correlated decoding is done per logical measurement and errors that affect such a measurement can occur at any space-time point in the backpropagation of the measured operator~\cite{CainFastCorrelatedDecoding2025, Serra-PeraltaDecodingTransversalClifford2026}.
While efficient decoding strategies have been developed, such as minimum-weight perfect matching that exploits the specific structure of certain codes~\cite{DennisTopologicalQuantumMemory2002, HiggottSparseBlossomCorrecting2025}, or heuristic algorithms such as belief propagation~\cite{PoulinIterativeDecodingSparse2008}, decoding is generally considered a problem that scales badly with problem size~\cite{HsiehNPhardnessDecodingQuantum2011, KuoHardnessDecodingQuantum2012, IyerHardnessDecodingQuantum2015}.
It thus remains an open question whether correlated decoding could become a bottleneck in practical implementations of large-scale fault-tolerant computations.

Here, we address this question by analyzing the complexity of the decoding problem in a setting motivated by recent experimental and theoretical developments: random circuits that combine single- and two-qubit Clifford operations with magic state injection.
A representative example, in which the two-qubit Clifford gates are arranged in a brickwork geometry, is illustrated in \cref{fig:circuit-model}.
We investigate how the decoding complexity, as measured by the size of decoding graphs, scales with the number of computational qubits and with circuit depth.
Our main finding is that the complexity remains bounded for a finite space-time density $\kappa$ of injected magic.

\begin{figure}[t]
    \centering
    \includegraphics[width=\linewidth]{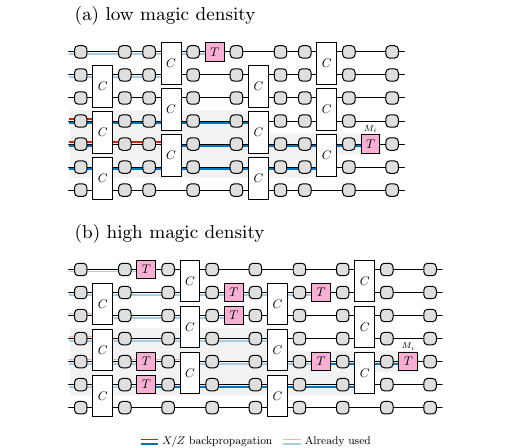}
    \caption{%
        Origin of bounded decoding complexity.
        The grey shading marks the space-time sensitivity region of the logical Pauli measurement $M_i$ associated with a magic state injection, obtained by propagating the measurement operator backward through the circuit.
        Red and blue lines indicate the $X$- and $Z$-Pauli content of the backpropagation, with faded lines corresponding to portions that have been already processed when decoding earlier measurements.
        Compared with (a) low magic state injection density, (b) a higher density $\kappa$ leads to more frequent mid-circuit measurements and leaves less new syndrome information to be processed per measurement.
    }
    \label{fig:sensitivity-regions}
\end{figure}

The basic mechanism underlying our result is illustrated schematically in \cref{fig:sensitivity-regions}.
To decode a logical Pauli measurement, it is propagated backward through the preceding Clifford circuit to identify the space-time sensitivity region in which errors can affect the measurement outcome.
When measurements are decoded sequentially, syndrome information that has already been processed for earlier measurements need not be revisited.
Consequently, only the previously unprocessed part of the sensitivity region contributes to the next decoding problem.
Magic state injection generates mid-circuit measurements throughout the computation, so that at a finite density $\kappa$ of injected magic states, these previously processed regions continually restrict the amount of new syndrome information that must be considered.
We find that the typical size of the resulting decoding graphs approaches a finite stationary value that is independent of both the number of computational qubits and the circuit depth, suggesting that scalable correlated decoding may be achievable in practice.
Moreover, the amount of new information to be processed decreases with increasing $\kappa$, as anticipated by the comparison in \cref{fig:sensitivity-regions}.
Interestingly, there is substantial freedom in choosing specific products of logical measurements to decode.
We find that the resulting decoding complexity is remarkably insensitive to this choice.

The paper is organized as follows.
We begin in \cref{sec:space-time-corr} with a review of space-time correlated decoding for fault-tolerant logical circuits composed of Clifford gates supplemented by magic states.
A novel justification of the sampling strategy underlying correlated decoding is given in \cref{sec:sampling}.
\Cref{sec:compl-rand-circ} is devoted to the decoding complexity in random circuits, with particular emphasis on the brickwork geometry shown in \cref{fig:circuit-model}.
Additional numerical results for other circuit layouts are presented in \cref{sec:other-layouts}, and \cref{sec:scal-form-stat} provides a detailed analysis of the scaling properties of the stationary decoding-volume distributions.
Finally, in \cref{sec:outlook}, we discuss the implications of our results for large-scale fault-tolerant quantum computation and outline directions for future research.

\section{Space-time correlated decoding in transversal circuits}
\label{sec:space-time-corr}

This section reviews the sampling and decoding strategy underlying space-time correlated decoding.
Rather than decoding every logical measurement individually, the method introduced in Refs.~\cite{CainFastCorrelatedDecoding2025, Serra-PeraltaDecodingTransversalClifford2026} identifies a set of \emph{reliable measurement products} whose outcomes suffice to reconstruct the full distribution of logical measurement outcomes
\footnote{%
    Throughout, we use the terms reliable and unreliable for measurements and measurement products, following Ref.~\cite{CainFastCorrelatedDecoding2025}; unreliable measurements are equivalent to the fragile measurements of Ref.~\cite{Serra-PeraltaDecodingTransversalClifford2026}.
}.
We explain first how these reliable products are constructed and then how they are decoded efficiently by exploiting the space-time structure of their backpropagation through the Clifford circuit.
This is complemented by a novel rigorous justification of the sampling strategy underlying correlated decoding presented in \cref{sec:sampling}.

\subsection{Constructing reliable measurement products}
\label{sec:reliable}

Consider a quantum computation implemented by an adaptive Clifford circuit acting on $N$ computational logical qubits and additional logical auxiliary qubits used for magic state injection, as illustrated in \cref{fig:circuit-model}.
The computational qubits are initialized in $+1$ eigenstates of the logical $X$ or $Z$ operators, denoted by $\ket{+}_L$ and $\ket{0}_L$, respectively, while the auxiliary qubits are prepared in logical magic states, $\ket{T}_L = T \ket{+}_L$.

The computation proceeds through a sequence of Clifford blocks $U_i$ and single-qubit Pauli measurements $M_i$, where $i \in \{ 1, \dotsc, N_M \}$.
We denote the corresponding measurement outcomes by $m_i = \pm 1$.
Each Clifford block may depend on the outcomes of previous measurements.
This occurs, in particular, in the feedforward required for magic state injection.
In the circuit of \cref{fig:circuit-model}, the Clifford blocks $U_i$ consist of two-qubit Clifford gates $C$, CNOT gates, and conditional single-qubit $S$ gates.

The output of the computation is encoded in the joint probability distribution $p(\mathbf{m})$ of measurement outcomes $\mathbf{m} = \left( m_1,\dotsc,m_{N_M} \right)$.
Space-time correlated decoding provides an efficient fault-tolerant method for reconstructing the distribution $p(\mathbf m)$ by sampling realizations of an encoded circuit~\cite{ZhouLowOverheadTransversalFault2025, CainFastCorrelatedDecoding2025, Serra-PeraltaDecodingTransversalClifford2026}.
Here, we describe the sampling strategy underlying  space-time correlated decoding.
The decoding procedure itself is discussed in Sec.~\ref{sec:decoding}.

In principle, sampling from $p(\mathbf{m})$ could be done straightforwardly by executing the Clifford blocks $U_i$ and measurements $M_i$ sequentially, thereby directly obtaining the complete sequence of measurement outcomes, $\mathbf{m} = \left( m_1, \dotsc, m_{N_M} \right)$.
However, determining $\mathbf{m}$ efficiently and fault-tolerantly in an encoded circuit is facilitated by adapting an alternative sampling strategy that focuses on measurements that are either \emph{reliable} or can be combined with earlier measurements into \emph{reliable products}; only these measurements have to be decoded.
By contrast, \emph{unreliable} measurements that cannot be combined to form reliable products need not be decoded, and their outcomes can be assigned uniformly at random.
Therefore, for each measurement $M_i$, we distinguish between the \emph{assigned measurement outcome} $m_i$ and the \emph{actual measurement outcome} $\widetilde{m}_i$.
The alternative sampling strategy, motivated by the examples below and developed rigorously in \cref{sec:sampling}, reconstructs the distribution $p(\mathbf{m})$ by sampling assigned measurement outcomes.

We classify measurements as reliable or unreliable according to the stabilizers of the initial state.
Each computational qubit $j\in\{1,\dotsc,N\}$ is associated with a logical Pauli stabilizer $G_j$, where $G_j=X$ for qubits initialized in $\ket{+}_L$ and $G_j=Z$ for qubits initialized in $\ket{0}_L$.
A measurement $M_i$ is called \emph{reliable} if the backpropagated measurement operator at state initialization,
\begin{equation}
    \label{eq:M-i-backpropagation}
    M_{i, 0} = U_1^{\dagger} \dotsb U_i^{\dagger} M_i U_i \dotsb U_1,
\end{equation}
commutes with every stabilizer of the initial state, and \emph{unreliable} otherwise~\cite{CainFastCorrelatedDecoding2025, Serra-PeraltaDecodingTransversalClifford2026}.
Logical auxiliary qubits prepared in magic states $\ket{T}_L$ do not possess Pauli stabilizers and therefore do not affect this classification.

In an adaptive circuit, the Clifford operations appearing in the backpropagation may depend on the outcomes of earlier measurements $M_{i'}$ with $i'<i$.
Consequently, whether a given measurement is reliable may itself depend on the measurement branch.
Nevertheless, the backpropagated operator can always be computed efficiently on a classical computer, since Clifford conjugation maps Pauli operators to Pauli operators.

To understand why we can randomly assign the outcomes of unreliable measurements that are not part of a reliable product, consider an unreliable measurement $M_i$ whose backpropagation $M_{i, 0}$ anticommutes with a stabilizer $G_j$.
Since $G_j$ stabilizes the initial state, applying it leaves the state unchanged.
Propagating the stabilizer through the circuit, however, flips the sign of the measurement operator, $M_i \to -M_i$.
Equivalently, the measurement statistics are identical whether one measures $M_i$ or $-M_i$, implying that the measurement outcomes $\widetilde{m}_i=\pm1$ occur with equal probability.
A simple example is the circuit
\begin{equation}
    \label{eq:X-preparation-Z-measurement}
    \scalebox{0.8}{%
        \begin{quantikz}
            \lstick{\textcolor{paulired}{$\ket{+}_L$}}\zband{1}{3}
            & & \meter{M = Z} & \setwiretype{n}
        \end{quantikz}%
    },
\end{equation}
where the stabilizer $G=X$ flips the measurement operator $M=Z$.
This can be seen graphically as the backpropagation of the $Z$ measurement (blue) terminates on an $X$ eigenstate (red).
Instead of decoding the actual outcome $\widetilde{m}_i$, we may thus simply assign a measurement outcome $m_i = \pm 1$ uniformly at random.

\begin{figure*}[t]
    \centering
    \includegraphics[width=\textwidth]{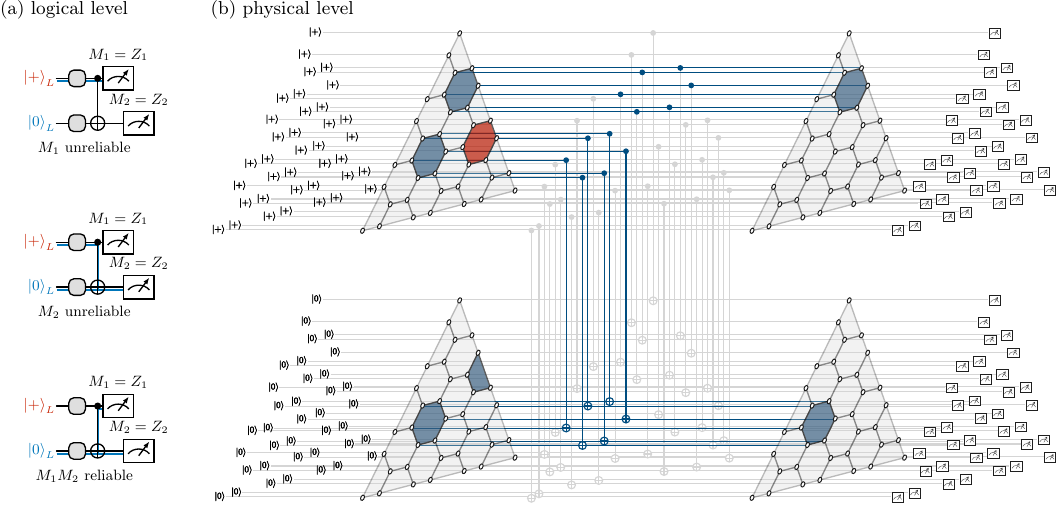}
    \phantomsubfloat{\label{fig:bell-logical}}
    \phantomsubfloat{\label{fig:bell-physical}}
    \caption{%
        Example of the construction of the decoding graph for Bell-state preparation.
        (a) At the logical level, the backpropagation of each individual measurement anticommutes with the initialization stabilizer $G_1=X_1$ of the first qubit and is therefore unreliable.
        The backpropagation of their product, by contrast, commutes with all initialization stabilizers, so that the product is reliable.
        (b) Construction of the corresponding decoding graph for two $\llbracket 37,1,7\rrbracket$ color-code logical qubits.
        Stabilizer measurements are related across circuit layers by the same Clifford-conjugation rules as logical Pauli operators (highlighted lines; only representative examples are shown).
        The construction of the checks is described in the main text and in \cref{eq:R-2-checks}.
        At the final time slice, products of the physical $Z$-basis measurement outcomes determine $Z$-type stabilizer outcomes, which form checks with the corresponding preceding stabilizer measurement (top) or with preceding stabilizer measurements on both logical qubits (bottom).
        The first measurements of $X$-type ($Z$-type) stabilizers form stand-alone checks for qubits initialized in $|+\rangle_L$ ($|0\rangle_L$), as illustrated at the top (bottom).
    }
    \label{fig:bell}
\end{figure*}

Reliable measurements, in contrast, cannot be treated in this way.
For example, replacing the $Z$ measurement in \cref{eq:X-preparation-Z-measurement} by an $X$ measurement yields the deterministic outcome $+1$.
The same is true whenever an unreliable measurement can be combined with earlier measurements to form a reliable product, indicating that correlations exist between their outcomes.
Such correlations arise when a stabilizer anticommutes not only with $M_i$ but also with previous measurements.
As an example, consider the circuit shown in \cref{fig:bell-logical} which prepares a Bell state between logical qubits $j=1$ and $2$ and measures both qubits in the $Z$ basis.
The joint distribution of measurement outcomes is $p(\widetilde{\mathbf{m}}) = \frac{1}{2} \delta_{\widetilde{m}_1, \widetilde{m}_2}$.
Both measurements being \emph{individually} unreliable is reflected in the marginal distributions, $p(\widetilde{m}_1) = p(\widetilde{m}_2) = 1/2$, meaning that the outcomes $\widetilde{m}_i = \pm 1$ occur with equal probability.
However, both measurements are flipped by the same stabilizer $G_1=X_1$ of the first logical qubit.
In the figure, this can be seen from the backpropagations of both measurements terminating as $Z_1$ (blue) on the initial $X$ eigenstate of the first qubit.
Consequently, when the measurements are performed sequentially, only the first measurement has equiprobable outcomes; owing to the Bell-state correlations, the outcome of the second measurement is deterministically equal to that of the first.
That is, although the second measurement is unreliable by itself, its product with the first measurement is reliable.
Graphically, the backpropagation of the product has support only on the second qubit, and its Pauli type $Z$ matches the initialization (both blue).
When sampling assigned outcomes $m_i$ instead of actual outcomes $\widetilde{m}_i$, this means that only $m_1 = \pm 1$ can be assigned uniformly at random, whereas the second measurement outcome has to be assigned such that also the assigned values obey Bell-state correlations.
Put differently, the reliable product has to be the same when expressed in terms of assigned or actual outcomes, $m_1 m_2 = \widetilde{m}_1 \widetilde{m}_2$.
This ensures that $\mathbf{m}$ and $\widetilde{\mathbf{m}}$ follow the same distribution.

While in this example the outcome for the reliable product was deterministic, this does not have to be the case.
In particular, the outcomes of reliable measurements and reliable products are probabilistic whenever their backpropagated operators have support on logical auxiliary qubits prepared in magic states $\ket{T}_L$.

In practice, reliable products are constructed sequentially as measurements are encountered during circuit execution.
To determine whether a reliable product containing a given measurement $M_i$ exists, we define the commutation matrix $C_{i,j} \in \mathbb{F}_2$ to equal $1$ whenever the backpropagation~\eqref{eq:M-i-backpropagation} of $M_i$ anticommutes with $G_j$, and $0$ otherwise.
When the measurement $M_i$ is performed, the corresponding row of the commutation matrix, $\mathbf{b}_i=(C_{i,j})_{j=1}^N$, is computed by backpropagating $M_i$ through the preceding Clifford circuit and checking its commutation with each initial stabilizer.
To construct a reliable product containing $M_i$, we write $R_i=\prod_{i'=1}^{N_M} M_{i'}^{V_{i',i}}$ with $V_{i',i} \in \mathbb{F}_2$.
Since the product must include $M_i$, but may involve only preceding measurements, its indicator vector takes the form $\mathbf{v}_i=\left(V_{i',i}\right)_{i'=1}^{N_M} =\left(\mathbf{x}_i,1,0,\dotsc,0\right)$, where $\mathbf{x}_i$ has $i-1$ components.
The condition for $R_i$ to be reliable is
\begin{equation}
    \label{eq:reliability-condition}
    \mathbf{v}_i C = 0 \quad \Longleftrightarrow \quad
    \mathbf{x}_i B_i = \mathbf{b}_i \pmod{2},
\end{equation}
where $B_i$ denotes the submatrix of $C$ consisting of its first $i-1$ rows.
The measurement $M_i$ is therefore reliable precisely when $\mathbf{b}_i = 0$ and Eq.~\eqref{eq:reliability-condition} is solved by $\mathbf{x}_i = 0$.
Otherwise, it can be incorporated into a reliable product if and only if Eq.~\eqref{eq:reliability-condition} admits a nontrivial solution.
In deep circuits with $i\gg N$, this solution is typically not unique.
We discuss different choices of reliable products below.

Maintaining a row-echelon form of $C$ throughout the circuit allows Eq.~\eqref{eq:reliability-condition} to be solved in $O(N^2)$ operations per measurement, where $N$ is the number of computational qubits and hence the number of initial stabilizers.
The total computational cost therefore scales as $O(N_MN^2)$, which is polynomial in the circuit size.

Having established how measurements are classified as reliable or unreliable, and how reliable products are constructed when they exist, we can now describe the sampling procedure used to reconstruct the joint distribution of measurement outcomes $p(\mathbf{m})$.

When $M_i$ is encountered during circuit execution, we first determine whether it is reliable or, if it is unreliable, whether it can be combined with previous measurements into a reliable product.
The latter is possible only  when Eq.~\eqref{eq:reliability-condition} admits a solution.
In that case, we write the corresponding reliable product as $R_i = \prod_{i'=1}^i M_{i'}^{v_{i'}}$, with $v_i=1$, and decode its outcome $r_i$.
Since all preceding measurements contained in $R_i$ have already been assigned outcomes, we assign an outcome $m_i$ such that $r_i = \prod_{i'=1}^i m_{i'}^{v_{i'}} = \prod_{i'=1}^i \widetilde{m}_{i'}^{v_{i'}}$.
The equality between the products of assigned and actual outcomes is discussed above for the example of Bell-state correlations and shown in full generality in \cref{sec:sampling}.
The special case in which $M_i$ is reliable by itself corresponds to $v_{i'}=0$ for all $i' \neq i$, so that the assigned outcome equals the actual one, $r_i = m_i = \widetilde{m}_i$.
If no reliable product containing $M_i$ exists, we do not decode $M_i$ and instead assign the outcome $m_i = \pm 1$ uniformly at random.

During circuit execution, Clifford operations are conditioned on the assigned rather than the actual measurement outcomes.
As explained in \cref{sec:sampling}, even if there is a mismatch between these values, sampling assigned outcomes correctly reproduces the distribution $p(\mathbf{m})$.

This sampling procedure forms the basis of space-time correlated decoding.
It therefore remains to specify how the outcomes of reliable products are decoded.

\subsection{Decoding reliable measurement products}
\label{sec:decoding}

We assume the computational and auxiliary qubits to be encoded in blocks of an $\llbracket n,k,d \rrbracket$ Calderbank--Shor--Steane (CSS) code that admits a transversal or fold-transversal implementation of a generating set of the Clifford group, such as $\{H,S,\cx\}$, where $H$ and $S$ denote the Hadamard and phase gates, respectively.
Examples include the two-dimensional color code, for which these Clifford generators can be realized transversally~\cite{BombinTopologicalQuantumDistillation2006}, and the surface code~\cite{KitaevFaulttolerantQuantumComputation2003, FowlerSurfaceCodesPractical2012}, where they can be realized fold-transversally~\cite{MoussaTransversalCliffordGates2016}.
Under such (fold-)transversal Clifford operations, code stabilizers are mapped to code stabilizers, with their Pauli content transforming according to the same Clifford conjugation rules as logical Pauli operators.
This property underlies the construction of the decoding graph described below.

Once a reliable product $R_i$ is identified, the decoder must determine its value from noisy physical syndrome data.
The corresponding decoding problem is obtained by backpropagating $R_i$ layer by layer through the circuit.
Its backpropagation to a layer $i' < i$ is defined as
\begin{equation}
    \label{eq:R-i-backpropagation}
    R_{i, i'} = U_{i' + 1}^{\dagger} \dotsb U_i^{\dagger} R_i U_i \dotsb U_{i' + 1}.
\end{equation}
At each layer $i'$, the support of $R_{i,i'}$ identifies the logical qubits on which an error can influence the value of $R_i$.
Taken over all layers $i' \in \{ 0, \dotsc, i\}$, these supports define a space-time sensitivity region for the measurement product.
A physical error within this region affects the value of $R_i$ precisely when it anticommutes with the corresponding backpropagated operator $R_{i,i'}$.
The Pauli type of $R_{i,i'}$ on each logical qubit therefore determines which code stabilizers are sensitive to the relevant errors, and hence which syndrome measurements must be included in the decoding problem.

The decoding graph for $R_i$ is constructed directly from this space-time sensitivity region.
At each layer, one includes the stabilizer measurements of the Pauli type selected by the backpropagated operator $R_{i,i'}$.
For notational simplicity, the backpropagation layers in Eq.~\eqref{eq:R-i-backpropagation} are defined by the blocks $U_i$ between successive logical measurements.
In practice, however, a block $U_i$ may have to be decomposed into individual transversal gate layers, with a single round of syndrome extraction inserted between them~\cite{ZhouLowOverheadTransversalFault2025, CainCorrelatedDecodingLogical2024, CainFastCorrelatedDecoding2025, Serra-PeraltaDecodingTransversalClifford2026}.
Each check is obtained by comparing a stabilizer measurement with the stabilizer, or product of stabilizers, to which it backpropagates through the preceding logical gate.
Consequently, a check may involve more than two stabilizer-measurement outcomes.
This occurs, for example, when a $Z$ stabilizer spreads onto another logical qubit under a transversal CNOT gate, or when an $X$ stabilizer is mapped to a product of $X$- and $Z$-type stabilizers by an $S$ gate.

We have so far described checks in the space-time bulk of the circuit.
Let us now turn to the temporal boundaries associated with logical-state preparation and measurement.
Computational qubits are initialized transversally in $\ket{+}^{\otimes n}$ or $\ket{0}^{\otimes n}$, corresponding to the logical states $\ket{+}_L$ and $\ket{0}_L$, respectively.
The first measurements of the corresponding $X$- or $Z$-type stabilizers therefore have deterministic outcomes and form stand-alone checks.
By contrast, the complementary stabilizers are initially undetermined: their first measurement outcomes are random and, in the presence of measurement errors, would require $O(d)$ rounds of syndrome extraction to determine reliably.

Auxiliary qubits prepared in logical magic states $\ket{T}_L$ are treated differently.
We assume these states to be prepared fault-tolerantly with well-defined eigenvalues of all code stabilizers.

Crucially, the backpropagation of a reliable product cannot terminate on an undetermined initial stabilizer.
If it did, then on at least one corresponding input qubit the backpropagated operator would contain the Pauli operator conjugate to the known initialization stabilizer.
It would therefore anticommute with that stabilizer, in contradiction with the reliability condition~\eqref{eq:reliability-condition}.
Hence, the decoding graph of a reliable product terminates only on stabilizers whose initial values are known~\cite{CainFastCorrelatedDecoding2025, Serra-PeraltaDecodingTransversalClifford2026}.

The final temporal boundary is set by logical measurement.
Logical measurements are performed transversally.
The code stabilizers of the same Pauli type as the logical measurement can therefore be inferred from products of the individual data-qubit measurement outcomes.
These are precisely the stabilizers sensitive to errors that can affect the logical measurement and are incorporated into the decoding problem together with the preceding syndrome measurements.

In \cref{fig:bell-physical}, we illustrate the construction of the decoding graph for Bell-state preparation, using the two-dimensional $\llbracket 37,1,7\rrbracket$ color code.
The two logical qubits, $j=1,2$, are initialized in $\ket{+}_L$ and $\ket{0}_L$ with a single round of syndrome extraction.
Syndrome information is obtained at two time slices, $i=1,2$, immediately after initialization and from the final transversal measurements.
We denote the outcome associated with stabilizer $l$ of Pauli type $\mu\in\{X, Z\}$ on logical qubit $j$ at time slice $i$ by $S_{j,l,i}^{\mu}$, with $l\in\{1,\dotsc,18\}$.
As discussed in \cref{sec:reliable}, both measurements $M_1 = Z_1$ and $M_2 = Z_2$ are individually unreliable, but the product $R_2 = Z_1 Z_2$ is reliable.
Its backpropagation is illustrated on the logical level in \cref{fig:bell-logical}.
Checks are constructed by backpropagation of stabilizers as illustrated in \cref{fig:bell-physical}, which gives
\begin{equation}
    \label{eq:R-2-checks}
    \begin{gathered}
        C_{1,l,2}^Z
        =S_{1,l,2}^ZS_{1,l,1}^Z,
        \quad
        C_{2,l,2}^Z
        =S_{2,l,2}^ZS_{1,l,1}^ZS_{2,l,1}^Z,
        \\
        C_{2,l,1}^Z=S_{2,l,1}^Z,
    \end{gathered}
\end{equation}
where the latter is a stand-alone check associated with the known initial stabilizer value.

Constructing checks by backpropagating stabilizers ensures that the syndromes of single data and measurement errors remain localized in space-time.
Consequently, data and measurement errors correspond to space-like and time-like hyperedges in the decoding graph, respectively.
Space-like hyperedges connect check nodes within the same time slice, whereas time-like hyperedges connect check nodes in successive time slices.

The weight of space-like hyperedges depends on the underlying code.
For the surface code, they have weight at most two, whereas for the color code they can have weight three.
By contrast, measurement errors give rise to time-like hyperedges that, in principle, connect up to three checks across a transversal logical gate.
However, irrespective of the chosen reliable product, only two of these checks appear in its decoding graph~\cite{CainFastCorrelatedDecoding2025, Serra-PeraltaDecodingTransversalClifford2026}.
Consequently, every time-like hyperedge reduces to an ordinary edge.
In the example shown in \cref{fig:bell}, a measurement error in $S_{2, l, 1}^z$ gives rise to a weight-two time-like edge connecting  checks $C_{2,l,2}^Z$ and $C_{2,l,1}^Z$ in \cref{eq:R-2-checks}.
If, in addition, the space-like hyperedges have weight at most two, as in the surface code, the resulting space-time decoding graph can be decoded efficiently using minimum-weight perfect matching.

Once the errors have been identified by the decoder, the corrected value of the reliable product is obtained by determining whether the decoded error anticommutes with the backpropagated measurement operator.

The computational cost of decoding a reliable product is determined by the number of checks in its decoding graph, which is set by the space-time extent of its backpropagation.
Decoding each reliable product independently would therefore repeatedly process the same regions of the circuit.
Moreover, because backpropagations typically extend all the way to the initialization of the logical qubits, the decoding graphs grow with circuit depth.

To avoid this overhead, reliable products are decoded in temporal order, committing the corresponding corrections after each decoding step~\cite{CainFastCorrelatedDecoding2025}.
Once a check has been decoded and its correction committed, it is never revisited when decoding subsequent products.
Consequently, each reliable product is decoded only within the previously unexplored portion of its backpropagation region, substantially reducing the effective decoding volume.

A subtle issue arises from weight-three time-like hyperedges associated with measurement errors.
As discussed above, for any choice of reliable product, only two of the three checks incident on such a hyperedge are included in the corresponding decoding graph.
Erroneously committing this hyperedge can therefore flip the third check, creating a \emph{virtual defect} in a subsequent decoding graph.
This becomes problematic if an odd number of such defects occurs within a connected space-time region, since at least one must then be matched to a spatial or temporal boundary and can thereby induce a logical error.
Such situations can arise in circuits containing time-like loops~\cite{CainFastCorrelatedDecoding2025, Serra-PeraltaDecodingTransversalClifford2026}.
Suitable circuit identities can sometimes be used to eliminate these loops.
When this is not possible, the problem can instead be resolved by introducing \emph{virtual edges}, corresponding to closed space-time loops within already decoded regions, which allow virtual defects in disconnected regions of subsequent decoding graphs to be matched consistently~\cite{CainFastCorrelatedDecoding2025}.

\section{Complexity in random circuits}
\label{sec:compl-rand-circ}

Having introduced space-time correlated decoding, we now analyze its computational complexity in a generic setting.
Previous work presented full-scale numerical simulations of the execution and decoding of transversal Clifford circuits encoded in the surface code~\cite{Serra-PeraltaDecodingTransversalClifford2026, CainFastCorrelatedDecoding2025}.
Circuit execution can be simulated efficiently with \texttt{stim}~\cite{GidneyStimFastStabilizer2021}, while minimum-weight perfect matching decoding is performed using \texttt{pymatching}~\cite{HiggottSparseBlossomCorrecting2025}.
Extending such full-scale simulations to circuits with magic state injection, however, is substantially more challenging, particularly when the circuits are deep and act on many computational qubits.
Consequently, the computational complexity of space-time correlated decoding for such circuits remains largely unexplored.

In particular, we seek to understand how the decoding complexity scales with the number of computational qubits and the circuit depth.
If the decoding cost were to grow with either, classical processing could become a bottleneck for real-time circuit execution.
One approach to controlling this growth is windowed decoding, which explicitly restricts decoding to finite space-time regions~\cite{ZhangScalableConstantTimeLogical2025, TurnerScalableDecodingProtocols2026, Serra-PeraltaDecodingTransversalClifford2026}.
Transversal operations, however, correlate syndrome information across logical qubits, making it challenging to define a decoding problem that preserves fault tolerance.
Here, we investigate a complementary mechanism in which finite decoding regions arise dynamically from the sequential decoding of reliable measurement products and the commitment of previous corrections.
For a broad class of circuits, the resulting decoding complexity can be characterized without simulating the full encoded dynamics.
As a proxy for the cost of decoding a reliable product, we use the size of its decoding graph, measured by the number of checks it contains, and exclude checks that have already been processed when decoding earlier reliable products.
As we show below, the resulting circuit-dependent decoding volume remains bounded: it grows neither with the number of computational qubits nor with the circuit depth.

\subsection{Circuit layout and complexity measure}
\label{sec:circuit-layout}

As a representative example, we consider the circuit layout shown in \cref{fig:circuit-model}.
The circuit comprises random two-qubit Clifford blocks arranged in a brickwork geometry and interspersed with $T$-gate injections occurring independently with probability $\kappa$.
Such Clifford blocks can be implemented transversally with the two-dimensional color code~\cite{BombinTopologicalQuantumDistillation2006}.
The qualitative conclusions of our analysis, however, are independent of the particular circuit layout.
In \cref{sec:other-layouts}, we present additional results for circuits with all-to-all connectivity and Clifford gates drawn from the set $\{H, S, \cx\}$, which admit fold-transversal implementations with surface-code logical qubits~\cite{MoussaTransversalCliffordGates2016, BreuckmannFoldTransversalCliffordGates2024}.

In general, circuits containing multiple layers of two-qubit gates give rise to time-like loops, which can lead to virtual defects~\cite{CainFastCorrelatedDecoding2025, Serra-PeraltaDecodingTransversalClifford2026}.
As discussed at the end of \cref{sec:decoding}, loops whose structure is known in advance can sometimes be simplified or eliminated using suitable circuit identities, with the corresponding preprocessing performed offline. In adaptive circuits, however, the loop structure may depend on whether conditional gates are executed, so that the appropriate simplifications may have to be determined for the realized measurement branch during circuit execution. Any loops that remain require the corresponding virtual edges and their weights to be constructed online, introducing additional classical overhead.
Since the size of the decoding graph remains bounded, however, we expect this overhead to remain bounded as well.

Although magic state injection introduces adaptive gates, it is nevertheless unnecessary to simulate the quantum state in order to characterize the decoding complexity.
Instead, it suffices to sample the sequence of logical gates appearing in the circuit.
To see this, recall that each $T$ gate is implemented by injecting a logical magic state $\ket{T}_L = T\ket{+}_L$.
The injection circuit, which is shown in \cref{fig:circuit-model}, consists of a CNOT gate from the computational qubit to the auxiliary magic state qubit, followed by a logical $Z$ measurement of the auxiliary qubit and a logical $S$ gate conditioned on the measurement outcome.
Depending on the circuit, the conditioning measurement may itself be unreliable or belong to a reliable product.
In either case, however, the backpropagated measurement operator terminates as $Z$ on the auxiliary $\ket{T}_L$ state, implying that the two measurement outcomes occur with equal probability.
Consequently, the conditional $S$ gate is applied with probability $1/2$.
The sequence of logical gates can therefore be generated simply by sampling each conditional $S$ gate independently with probability $1/2$, without simulating the underlying quantum state.

As a proxy for the complexity of decoding a reliable product $R_i$, we use the normalized check count $N_C$ of its decoding graph after excluding checks that have already been processed.
More precisely, suppose that, upon backpropagating $R_i$ to an earlier circuit step $i' < i$, its restriction to logical qubit $j$ is $X_j$ or $Z_j$.
The corresponding set of $X$- or $Z$-type checks is then included in the decoding graph of $R_i$, contributing $1$ to $N_C$.
If the restriction is $Y_j$, both sets of checks are included, contributing $2$.
Checks that have already appeared in previously decoded graphs are excluded.

Strictly speaking, $N_C$ is not the literal number of stabilizer checks in the decoding graph.
Rather, the actual number depends on the code used to encode each logical qubit.
For codes with equal numbers of $X$- and $Z$-type stabilizers, including the unrotated surface code and self-dual codes such as the color code, the actual number of checks is obtained by multiplying $N_C$ by the number of stabilizers of either type.
For a distance-$d$ unrotated surface code and color code, this factor is $d \left( d-1 \right)$ and $3 \left( d^2-1 \right) \! /8$, respectively.
Thus, while the normalized check count $N_C$ captures the dependence of the decoding problem on the logical circuit, its absolute size retains the dependence on the underlying code through the number of stabilizer checks per logical block.
In particular, this number grows with the code distance $d$, even though $N_C$ remains bounded with increasing circuit width and depth.

\subsection{Decoding complexity}
\label{sec:decoding-complexity}

\begin{figure}
    \centering
    \includegraphics[width=\linewidth]{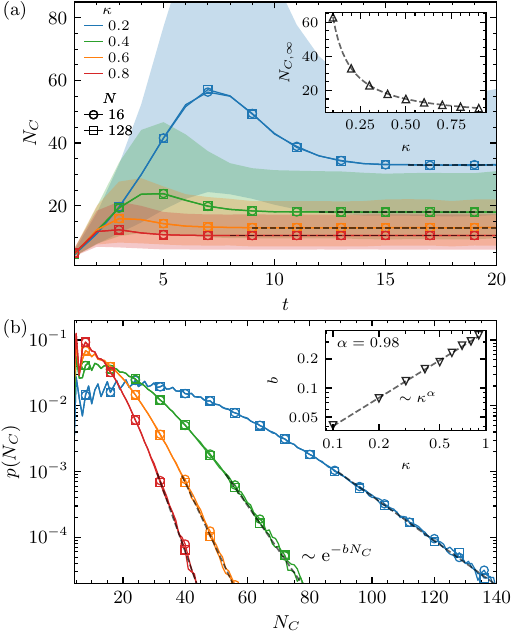}
    \phantomsubfloat{\label{fig:complexity-time}}
    \phantomsubfloat{\label{fig:complexity-distribution}}
    \caption{Decoding complexity in random circuits for different numbers of computational qubits $N$ and magic state injection probabilities $\kappa$.
        Colors indicate $\kappa$, while marker shapes distinguish $N=16$ and $128$.
        (a) Average normalized check count $N_C$ for mid-circuit measurements performed at circuit layer $t$.
        Black dashed lines indicate fits to the stationary regime.
        Inset: Stationary values $N_{C,\infty}$ extracted from these fits (markers), together with the analytical prediction in \cref{eq:N-C-infty} (dashed line).
        (b)~Probability distribution $p(N_C)$ in the stationary regime.
        The dashed lines indicate exponential fits to the large-$N_C$ tails.
        Inset: Fitted decay rate $b$ versus $\kappa$.
    }
    \label{fig:complexity}
\end{figure}

\Cref{fig:complexity-time} shows the normalized check count $N_C$ as a function of the circuit layer index $t$ for $N = 16$ and $128$ computational logical qubits and four values of the magic state injection probability $\kappa$.
Each circuit layer consists of a Clifford layer followed by a $T$-gate layer, as shown in \cref{fig:circuit-model}.
The data are averaged over $N_{\mathrm{real}} = 10^4$ and $10^3$ circuit realizations for $N = 16$ and $128$, respectively.
Since each circuit layer contains, on average, $\kappa N$ magic state injections, each data point is obtained from approximately $\kappa N N_{\mathrm{real}}$ decoding instances.
The shaded regions indicate the spread of the data and are defined by upper and lower standard deviations, computed separately from samples above and below the mean and normalized such that they reduce to the conventional standard deviation for a symmetric distribution.

We first consider the transient behavior at early circuit times.
During the initial stages of circuit execution, only a small number of previous measurements are available to form reliable products.
Consequently, the first $O(N)$ measurements are typically unreliable and are not decoded immediately.
Their associated checks therefore accumulate until they are eventually incorporated into the decoding graphs of subsequent reliable products, causing the average normalized check count $N_C$ to increase.
This growth continues until approximately $N$ unreliable measurements have accumulated, corresponding to a transient regime of duration $t\sim\kappa^{-1}$.
As reliable products subsequently become available, the accumulated backlog of checks is gradually depleted, producing the overshoot visible in Fig.~\ref{fig:complexity-time}.
The check count then relaxes towards a stationary regime in which its average becomes independent of circuit depth.
Increasing $\kappa$ shortens the transient and reduces the height of the overshoot because reliable products are generated more rapidly.

To understand the stationary regime analytically, note first that the number of unreliable measurements that cannot be combined with previous measurements is bounded by the number of computational qubits $N$.
This follows because one can choose a basis of stabilizers of the initial state such that every such measurement is flipped by exactly one basis element, and the number of basis stabilizers equals the number of computational qubits~\cite{CainFastCorrelatedDecoding2025}.
Once this bound is reached, every subsequent measurement can be incorporated into a reliable product.
A stationary regime is then established in which checks are generated and decoded at equal average rates.
Inspection of \cref{fig:circuit-model} shows that each circuit layer contributes $2N$ checks from the Clifford gates and another $4N$ checks from the $T$-gate layer, corresponding to one $X$ and one $Z$ check per computational qubit.
In addition, each magic state injection contributes three checks: one obtained by backpropagating the logical $Z$ measurement of the auxiliary qubit and two stand-alone checks associated with the logical $\ket{T}_L$ state, as discussed in Sec.~\ref{sec:decoding}.
Since the mean number of magic state injections per layer is $\kappa N$, the average number of newly generated checks per layer is $3 \left( 2 + \kappa \right) N$.
Although a given check may be contained in the decoding graphs of many subsequent reliable products, it is counted only the first time it appears.
Consequently, in the stationary regime, the average number of newly counted checks per layer equals the average number of newly generated checks.
Dividing by the mean number of reliable products created per layer, $\kappa N$, yields the stationary value
\begin{equation}
    \label{eq:N-C-infty}
    N_{C,\infty}=\frac{3 \left( 2 + \kappa \right)}{\kappa}.
\end{equation}
More generally, for all circuit architectures that reach such a stationary regime, this counting argument shows that the stationary normalized check count is given by the ratio of the average number of newly generated checks per circuit layer to the average number of reliable products created per layer.

The numerical data shown in \cref{fig:complexity-time} are in excellent agreement with this prediction.
In particular, the stationary value is independent of the number of computational qubits, as demonstrated by the collapse of the data for $N=16$ and $128$.
This follows because both the number of generated checks and the number of reliable products per layer are proportional to $N$.
The dependence of $N_{C,\infty}$ on $\kappa$ is shown in the inset of \cref{fig:complexity-time}, where Eq.~\eqref{eq:N-C-infty} is found to accurately describe the numerical results.
Notably, the decoding complexity per reliable product decreases with increasing magic state injection probability.

Having established that the average normalized check count remains bounded with both the number of computational qubits and the circuit depth, we next consider fluctuations around the stationary value.
The stationary distribution $p(N_C)$ is shown in \cref{fig:complexity-distribution}.
As for the mean, the results for $N=16$ and $128$ collapse, indicating that the full stationary distribution is likewise insensitive to the number of computational qubits.
At large $N_C$, the distribution exhibits an exponential tail, $p(N_C) \sim \mathrm{e}^{-b N_C}$.
The decay rate $b$ increases with the magic state injection density approximately as $\sim \kappa^{\alpha}$ where $\alpha = 0.98$; this behavior is shown in the inset of \cref{fig:complexity-distribution}.
Thus, increasing $\kappa$ suppresses not only the average decoding complexity but also the probability of large decoding instances.

The fitted exponent $\alpha = 0.98$ suggests an almost linear dependence $b \sim \kappa^{\alpha}$.
However, the detailed analysis of the stationary distribution presented in \cref{sec:scal-form-stat} favors scaling behavior with a nontrivial exponent $\alpha < 1$.
In particular, we find that for circuits composed of random two-qubit Clifford gates, the distribution obeys the one-parameter scaling form
\begin{equation}
  \label{eq:scaling-form}
  p(N_C) = \kappa^{\alpha} F \! \left[ \kappa^{\alpha} \left( N_C - N_{C, \infty} \right) \right].
\end{equation}
While the data hints that $\alpha$ and $F$ are independent of the circuit connectivity, we cannot determine conclusively if this is a universal behavior across random one-dimensional brickwork and all-to-all circuits.
By contrast, restricting the gate ensemble to a subset of the Clifford group leads to deviations from this scaling form.
Understanding the origin of this scaling behavior and the conditions under which it emerges remains an interesting direction for future research.

\subsection{Choice of reliable measurement products}
\label{sec:choice-product}

As discussed in \cref{sec:reliable}, the choice of a reliable product containing a given measurement $M_i$ is generally not unique.
We now investigate how this freedom in choosing $R_i$ affects the decoding complexity.

We first quantify how many different reliable products can be associated with a given measurement $M_i$.
Each such product corresponds to a different solution $\mathbf{x}_i$ of \cref{eq:reliability-condition}.
Given a particular solution, the entire solution space is obtained by adding vectors in the left nullspace of $B_i$.
The dimension of the solution space is therefore $i - 1 - \operatorname{rank}(B_i)$.
Hence, determining the number of possible reliable products reduces to determining the rank of $B_i$.
For a deep circuit with $i\gg N$, $B_i$ is a tall matrix: it has only $N$ columns, but each column contains $i-1$ entries.
We now argue that these columns are linearly independent with high probability, meaning that $\operatorname{rank}(B_i)=N$.

To see this, recall the definition of $B_i$.
Its row indexed by $i'<i$ describes the commutation of the backpropagated measurement $M_{i'}$ with the $N$ Pauli stabilizers $G_j$ of the initial computational state: the matrix element $(B_i)_{i',j}$ equals $1$ if the backpropagation of $M_{i'}$ anticommutes with $G_j$, and $0$ otherwise.
In a sufficiently deep random circuit, the restriction of a backpropagated measurement to a given computational qubit is approximately uniformly distributed over $\{I,X,Y,Z\}$.
Since $G_j$ is either $X_j$ or $Z_j$, half of these Pauli operators commute with $G_j$ and half anticommute.
The entries of $B_i$ can therefore be regarded, to a good approximation, as unbiased binary variables.
Accordingly, its $N$ columns behave approximately as long random binary strings.
For $i-1\gg N$, such strings are overwhelmingly likely to be linearly independent over $\mathbb{F}_2$, so that $\operatorname{rank}(B_i)=N$ with high probability.
The solution space then has dimension $i-1-N$ and contains $2^{i-1-N}$ solutions.
Thus, the number of possible reliable products grows exponentially with circuit depth.

To determine whether this freedom affects the decoding complexity, we compare four different prescriptions for selecting a solution of \cref{eq:reliability-condition}.
These prescriptions produce solutions with different Hamming weights and hence reliable products containing different numbers of measurements.
As a simple baseline, which we label Free, we solve the transposed system~\eqref{eq:reliability-condition} by Gaussian elimination and set all free variables to zero.
This is the prescription used for the results presented in the previous subsection and in \cref{fig:complexity}.
Since $B_i^{\top}$ has at most $N$ pivot columns, at most $N$ components of $\mathbf{x}_i$ can be nonzero.
Including the current measurement $M_i$, the Hamming weight $h_i$ of the full indicator vector $\mathbf{v}_i= \left(\mathbf{x}_i,1,0,\dotsc,0\right)$, and hence the number of measurements contained in $R_i$, therefore satisfies $h_i\leq N+1$.
We further consider a solution labeled Random, obtained by adding a uniformly sampled vector from the homogeneous solution space to the Free solution.
For $i\gg N$, such a random solution is approximately an unbiased binary vector of length $i-1$, subject only to $N$ linear constraints.
Approximately half of its components are therefore nonzero, giving $h_i \sim \left( i+1 \right) \! /2$ and thus a Hamming weight that grows linearly with $i$.
Finally, we consider a low-Hamming-weight solution obtained using belief propagation with ordered-statistics decoding (BP+OSD), as implemented in the Python package \texttt{ldpc}~\cite{RoffeLDPCTools2022, RoffeDecodingQuantumLowdensity2020}, and an exact minimum-Hamming-weight solution obtained using a MaxSAT solver from \texttt{pysat}~\cite{IgnatievPySATPythonToolkit2018}.

\begin{figure}
    \centering
    \includegraphics[width=\linewidth]{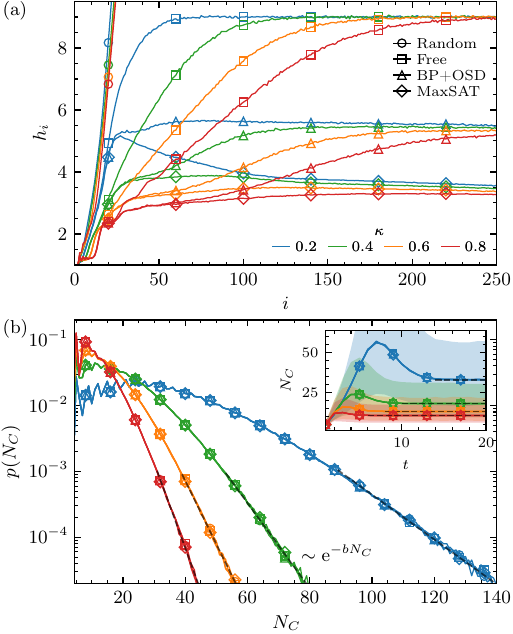}
    \phantomsubfloat{\label{fig:reliable-time}}
    \phantomsubfloat{\label{fig:reliable-distribution}}
    \caption{%
        Dependence of the decoding complexity on the choice of reliable measurement product for $N=16$.
        Colors indicate the magic state injection probability $\kappa$, while marker shapes distinguish the four methods used to select a solution of \cref{eq:reliability-condition}.
        (a) Hamming weight $h_i$ of the reliable product $R_i$ as a function of the measurement index $i$.
        Random denotes a random solution; Free the Gaussian-elimination solution with all free variables set to zero; BP+OSD a low-Hamming-weight solution; and MaxSAT the exact minimum-Hamming-weight solution.
        (b) Stationary distribution $p(N_C)$ of the normalized check count.
        Inset: Average $N_C$ versus circuit layer $t$.
        Despite the pronounced differences in $h_i$, the decoding-complexity statistics are indistinguishable within numerical resolution.
    }
    \label{fig:reliable}
\end{figure}

The Hamming weights obtained with these four prescriptions are compared in \cref{fig:reliable-time} for the same random-circuit ensemble considered above, with $N=16$.
For the Random choice, $h_i$ quickly leaves the range of the plot due to its approximately linear growth with the measurement index $i$.
By contrast, the other three prescriptions produce much smaller reliable products.
For Free, the boundedness follows directly from the analytical bound $h_i\leq N+1$ derived above.
BP+OSD typically yields lower Hamming weights than Free, while MaxSAT gives the exact minimum-Hamming-weight solution.
The four methods therefore generate reliable products whose numbers of constituent measurements differ substantially.

Despite these pronounced differences in the sizes of the reliable products, their decoding complexities are essentially indistinguishable.
\cref{fig:reliable-distribution} shows the stationary distribution of the normalized check count $N_C$, while the inset shows its average as a function of circuit layer $t$.
No discernible dependence on the prescription used to select the reliable product is observed.
Thus, minimizing the number of measurements contained in $R_i$ does not translate into a reduction of the effective decoding volume.
For the random circuits considered here, computationally expensive optimization procedures such as MaxSAT are therefore unnecessary: the Free solution obtained efficiently by Gaussian elimination is sufficient.

\section{Conclusions and outlook}
\label{sec:outlook}

Magic state injection is usually regarded as the most resource-demanding step in fault-tolerant quantum computation.
Yet, for space-time correlated decoding, our results reveal a counterbalancing effect: magic state injections introduce mid-circuit measurements that enable certain corrections to be committed, thereby reducing the complexity of decoding.
In random circuits composed of transversal or fold-transversal Clifford gates and magic state injections, these measurements continually break the decoding problem into pieces of finite size, provided the magic state density $\kappa$ remains finite.
The decoding volume associated with each reliable measurement product approaches a stationary average distribution that is independent of both the number of computational qubits and the circuit depth.
Its mean is determined by a simple balance between the rates at which checks are generated and consumed, as expressed by \cref{eq:N-C-infty}, while large decoding instances are exponentially suppressed.

These findings have direct implications for the classical cost associated with correlated decoding.
For $\kappa = 0.5$, for instance, the average decoding problem is comparable in size to a memory experiment for a single logical qubit over about fifteen syndrome-extraction rounds, independent of the width or depth of the computation.
As in the memory experiment, the absolute decoding cost retains a dependence on the underlying quantum code: it is obtained by multiplying the decoding volume by the number of relevant stabilizers per code block, which grows with the code distance.
While a memory experiment is restricted to a single logical qubit, bounded decoding regions thus emerge dynamically in our approach, even for computations involving arbitrarily many logical qubits.
This provides an attractive alternative to windowed decoding, which likewise aims to bound the real-time decoding problem in large-scale fault-tolerant computations.
Windowed decoding confines the decoding problem to a finite moving region of space-time~\cite{ZhangScalableConstantTimeLogical2025, TurnerScalableDecodingProtocols2026, Serra-PeraltaDecodingTransversalClifford2026}; for circuits with transversal operations, however, determining boundaries at which corrections can be safely committed while preserving fault tolerance is nontrivial.
By contrast, the approach considered here sequentially decodes reliable measurement products and commits the corresponding corrections, thereby removing syndrome information that has already been processed and dynamically determining the boundaries of the decoding regions.
Crucially, these commitments preserve fault tolerance by construction, without requiring additional prescriptions~\cite{TurnerScalableDecodingProtocols2026}.

Additionally, sequential decoding of reliable measurement products does introduce additional complexity into classical real-time control.
In particular, preparing each decoding problem requires identifying a reliable measurement product and constructing the corresponding decoding graph.
Interestingly, however, the freedom in choosing reliable measurement products does not appear to require optimization.
Although the number of reliable products available for a given measurement grows exponentially with circuit depth, the space-time volume to decode is effectively localized around the latest measurement, and the choice of the reliable product does not matter.
The prescriptions considered here yield indistinguishable statistics within our numerical resolution; for the random circuits studied here, the simple Gaussian-elimination solution is therefore sufficient.

Our results open several promising directions for future research.
Because the decoding volume is controlled by the local density of magic rather than by circuit size, it becomes a quantity that can be optimized at compile time: distributing magic state injections as uniformly as possible in space and time, within the constraints of a given algorithm, directly limits the decoding load.
Extending our analysis from random circuits to deep, structured circuits representative of concrete algorithms is therefore a natural next step toward a joint optimization of compilation and decoding.
The stationary distribution of decoding volumes, in turn, provides a basis for determining the computational resources required for real-time decoding, including the capacity to handle rare, unusually large decoding instances.
Finally, for matchable codes, it would be desirable to develop a systematic procedure for ensuring matchability of space-time decoding in circuits containing time-like loops.
As discussed in \cref{sec:decoding}, measurement errors in such circuits can generate virtual defects whose decoding requires the introduction of virtual edges.
While virtual edges arising from certain elementary gate sequences preserve matchability of the underlying code, this is generally no longer the case for longer and more complex circuits~\cite{CainFastCorrelatedDecoding2025}.
Whether and how such circuits can be transformed to restore matchability remains an open question.
Resolving this issue would unlock the use of fast matching decoders for real-time processing of the bounded decoding volumes identified here.

Taken together, our results indicate that the classical processing for transversal fault-tolerant architectures can scale alongside the quantum hardware, and that the magic states that make these architectures universal also help keep them decodable.

\begin{acknowledgments}
    We used OpenAI ChatGPT (GPT-5.6 Pro) and Anthropic Claude (Fable 5.1) to assist with refining and checking selected technical arguments, writing and debugging code used in our numerical analysis, identifying relevant literature, and editing the manuscript.
    We critically reviewed and revised all generated outputs and take full responsibility for the scientific content and conclusions.
    G.G.\ acknowledges support from the German Federal Ministry of Education and Research (BMBF) through the funded project ALMANAQC, grant number 13N17236 within the research program ``Quantum Systems.''
\end{acknowledgments}

\section*{Data availability}

The data that support the findings of this article are publicly available~\cite{SiebererDecodingData2026}.

\appendix

\section{Sampling assigned measurement outcomes}
\label{sec:sampling}

In \cref{sec:reliable}, we argued that sampling \emph{assigned} measurement outcomes is equivalent to sampling the actual measurement outcomes.
Here, we establish this result.
Since the argument is independent of the particular decoding method, we formulate the discussion at the logical level and, for simplicity, do not distinguish explicitly between logical and physical qubits.

Recall that, in the computational model considered here, a quantum computation is implemented by an adaptive Clifford circuit.
Formally, such a circuit consists of a sequence of unitary Clifford blocks $U_i$ and single-qubit Pauli $X$ and $Z$ measurements $M_i$ with outcomes $m_i = \pm 1$, where $i \in \{ 1, \dotsc, N_M \}$ labels gates and measurements in temporal order.
The adaptivity of the circuit implies that the gates may depend on previous measurement outcomes, i.e., $U_i = U_i(m_1, \dotsc, m_{i - 1})$.
We further assume that qubits that have been measured do not participate in the subsequent unitary evolution, so that $\left[ M_i, U_{i'} \right] = 0$ for all $i < i'$.
This assumption is satisfied, in particular, for auxiliary qubits used in magic state injection.

The outcome of the computation is encoded in the joint probability distribution of the $N_M$ measurement outcomes, denoted by $p(\mathbf{m})$, where $\mathbf{m} = \left( m_1, \dotsc, m_{N_M} \right)$.
This distribution can be obtained by sampling realizations of the circuit.
In each experimental run, the gate operations $U_i = U_i(m_1, \dotsc, m_{i - 1})$ and measurements $M_i$ are performed sequentially, with each gate $U_i$ conditioned on the actual outcomes $m_{i'}$ of preceding measurements.

The initial state $\ket{\psi_0}$ is a product state of logical computational qubits prepared in $\ket{+}$ or $\ket{0}$ states and auxiliary qubits prepared in magic $\ket{T}$ states where, as noted above, we omit the subscript $L$ indicating that these are logical states.
After $N_M$ Clifford blocks $U_i$ and measurements, $M_i$, the unnormalized state reads
\begin{equation}
    \label{eq:psi}
    \ket{\psi(\mathbf{m})} = P_{N_M} U_{N_M} \dotsb P_1 U_1 \ket{\psi_0},
\end{equation}
where the projection operators are given by
\begin{equation}
    \label{eq:projector}
    P_i = \frac{1}{2} \left( 1 + m_i M_i \right).
\end{equation}
The probability of obtaining the sequence of measurement outcomes $\mathbf{m}$ when sampling the circuit is given by the norm of the state:
\begin{equation}
    \label{eq:P-m}
    p(\mathbf{m}) = \norm{\ket{\psi(\mathbf{m})}}^2.
\end{equation}
This representation of the probability distribution forms the starting point for establishing the equivalence between sampling actual and assigned measurement outcomes.

We thus distinguish for each measurement $M_i$ between the \emph{assigned} outcome $m_i$ and the \emph{actual} outcome $\widetilde{m}_i$.
The procedure for sampling assigned measurement outcomes is as follows.
In step $i$ of the circuit execution, we first apply the gates $U_i = U_i(m_1, \dotsc, m_{i - 1})$, conditioned on the assigned values of the previous measurements.
To determine the assigned value $m_i$, we check whether $M_i$ is reliable or, if $M_i$ is unreliable, whether it is possible to form a reliable product containing $M_i$ and previous measurements.
As discussed in Sec.~\ref{sec:reliable}, this can be done by solving a linear system over $\mathbb{F}_2$.
We denote the set of such measurements by $\mathsf{R} \subseteq \{ 1, \dotsc, N_M \}$.
For $i \in \mathsf{R}$, we write the reliable Pauli product as $R_i = \prod_{i' = 1}^i M_{i'}^{V_{i', i}}$, where $V_{i', i} \in \mathbb{F}_2$ with $V_{i, i} = 1$ determines whether $M_{i'}$ is part of the reliable product.
As noted above, the case in which $M_i$ is reliable by itself is included as the special instance where $V_{i', i} = 0$ for $i \neq i'$.

We then measure the reliable product.
At this point, all previous measurements $M_{i'}$ with $i' \in \{ 1, \dotsc, i - 1 \}$ have already been performed.
Therefore, measuring $R_i$ reduces to measuring $M_i$ with actual outcome $\widetilde{m}_i$, corresponding to the outcome $r_i = \prod_{i' = 1}^i \widetilde{m}_{i'}^{V_{i', i}}$ for the measurement of $R_i$.
The assigned value $m_i$ is then chosen such that the value of the reliable product is the same when expressed in terms of assigned values,
\begin{equation}
    \label{eq:reliable-product-m-tilde-m}
    r_i = \prod_{i' = 1}^i m_{i'}^{V_{i', i}} = \prod_{i' = 1}^i \widetilde{m}_{i'}^{V_{i', i}},
\end{equation}
which implies
\begin{equation}
    \label{eq:m-from-tilde-m}
    m_i = r_i \prod_{i' = 1}^{i - 1} m_{i'}^{V_{i', i}} = \widetilde{m}_i \prod_{i' = 1}^{i - 1} \left( m_{i'} \widetilde{m}_{i'} \right)^{V_{i', i}}.
\end{equation}
If, instead, $M_i$ is unreliable and is not part of any reliable product, we assign an outcome $m_i = \pm 1$ uniformly at random.
We denote the set of these measurements by $\mathsf{U} = \{ 1, \dotsc, N_M \} \setminus \mathsf{R}$.

The resulting unnormalized state is
\begin{equation}
    \label{eq:tilde-psi}
    \ket{\widetilde{\psi}(\mathbf{m}, \widetilde{\mathbf{m}})} = \widetilde{P}_{N_M} U_{N_M} \dotsb \widetilde{P}_1 U_1 \ket{\psi_0}.
\end{equation}
This should be compared with Eq.~\eqref{eq:psi}.
There, both the gates $U_i$ and the projectors $P_i$, which describe the collapse of the state induced by the measurement $M_i$, are determined by the same outcomes $m_i$.
In Eq.~\eqref{eq:tilde-psi}, by contrast, the assigned values $m_i$ determine only the gates $U_i$, while the collapse of the state is governed by the actual measurement outcomes $\widetilde{m}_i$,
\begin{equation}
    \label{eq:tilde-projector}
    \widetilde{P}_i = \frac{1}{2} \left( 1 + \widetilde{m}_i M_i \right).
\end{equation}
Writing the state in this form already implicitly uses the fact that, for $i \in \mathsf{R}$, measuring $R_i$ with outcome $r_i$ is equivalent to measuring $M_i$ with outcome $\widetilde{m}_i$, since all previous measurements appearing in the product have already been performed.
The two outcomes are related by Eq.~\eqref{eq:reliable-product-m-tilde-m}.
To make this equivalence explicit, note that Eq.~\eqref{eq:reliable-product-m-tilde-m} can be rewritten as
\begin{equation}
    \widetilde{m}_i = r_i \prod_{i' = 1}^{i - 1} \widetilde{m}_{i'}^{V_{i', i}}.
\end{equation}
The projector~\eqref{eq:tilde-projector} can therefore be expressed as
\begin{equation}
    \label{eq:projector-reliable}
    \widetilde{P}_i = \frac{1}{2} \left[ 1 + r_i \left( \prod_{i' = 1}^{i - 1} \widetilde{m}_{i'}^{V_{i', i}} \right) M_i \right].
\end{equation}
In Eq.~\eqref{eq:tilde-psi}, the projectors corresponding to $i \in \mathsf{R}$ are multiplied from the right by projectors $\widetilde{P}_{i'}$ associated with earlier measurements, $i' \in \{ 1, \dotsc, i - 1 \}$.
Since measured qubits do not participate in the subsequent evolution, the projectors $\widetilde{P}_{i'}$ commute with all unitary gate operators to their left.
Using the identity $\widetilde{m}_{i'} \widetilde{P}_{i'} = M_{i'} \widetilde{P}_{i'}$ together with $V_{i, i} = 1$, we may therefore rewrite $\widetilde{P}_i$ as
\begin{equation}
    \label{eq:reliable-product-projector}
    \widetilde{P}_i = \frac{1}{2} \left( 1 + r_i R_i \right) = P_i.
\end{equation}
In this form, it becomes clear that the projectors $\widetilde{P}_i$ with $i \in \mathsf{R}$ enforce agreement between the actual and assigned outcomes of the reliable Pauli products $R_i$.
Consequently, the state~\eqref{eq:tilde-psi}, with the projectors written in this form, describes the logical state encountered in correlated decoding, where the products $R_i$ are obtained through transversal measurements of all logical qubits contained in the product.
The assigned outcomes are then completely determined by the randomly assigned values $m_i$ for measurements $i \in \mathsf{U}$ and the values of the reliable products $r_i$ for $i \in \mathsf{R}$, which can be decoded efficiently.

As in Eq.~\eqref{eq:P-m}, the probability of observing a sequence of assigned and actual measurement outcomes, $\mathbf{m}$ and $\widetilde{\mathbf{m}}$, respectively, is given by the norm of the state,
\begin{equation}
    \label{eq:tilde-P-m-tilde-m}
    \widetilde{p}(\mathbf{m}, \widetilde{\mathbf{m}}) = \frac{1}{2^{\abs{\mathsf{U}}}} \lVert \ket{\widetilde{\psi}(\mathbf{m}, \widetilde{\mathbf{m}})} \rVert^2.
\end{equation}
The prefactor accounts for the probability of selecting a particular sequence of randomly assigned outcomes for measurements $i \in \mathsf{U}$ that are unreliable and cannot be combined with previous measurements to form a reliable product.
We denote the number of such measurements by $\abs{\mathsf{U}}$.
As we now show, the distribution of assigned measurement outcomes coincides with the distribution of actual measurement outcomes given in Eq.~\eqref{eq:P-m},
\begin{equation}
    \label{eq:P-equals-tilde-P}
    p(\mathbf{m}) = 2^{\abs{\mathsf{U}}} \widetilde{p}(\mathbf{m}, \widetilde{\mathbf{m}}).
\end{equation}
On the right-hand side, $\mathbf{m}$ and $\widetilde{\mathbf{m}}$ are not completely independent.
For $i \in \mathsf{R}$, the assigned outcome $m_i$ is determined by the actual outcome $\widetilde{m}_i$ together with the previous assigned and actual outcomes through Eq.~\eqref{eq:m-from-tilde-m}.
By contrast, for $i \in \mathsf{U}$, the assigned outcome $m_i$ is chosen independently and uniformly at random, irrespective of the actual outcome $\widetilde{m}_i$.
We therefore need to show that the two probability distributions remain identical even when assigned and actual outcomes differ for an arbitrary subset $\mathsf{M} \subseteq \mathsf{U}$ of unreliable measurements.

We first establish Eq.~\eqref{eq:P-equals-tilde-P} for the case in which all randomly assigned outcomes $m_i$ with $i \in \mathsf{U}$ coincide with the corresponding actual outcomes $\widetilde{m}_i$, that is, when $\mathsf{M} = \emptyset$.
In this situation, the assigned and actual outcomes also agree for all $i \in \mathsf{R}$.
This can be shown inductively.
Consider the smallest $i \in \mathsf{R}$.
In Eq.~\eqref{eq:m-from-tilde-m}, the product on the right-hand side involves only indices $i' \in \mathsf{U}$, for which the assigned and actual outcomes coincide by assumption.
It follows immediately that $m_i = \widetilde{m}_i$.
The same argument can then be applied successively to all remaining $i \in \mathsf{R}$ in increasing order.

Since $\mathbf{m} = \widetilde{\mathbf{m}}$, the projectors~\eqref{eq:projector} and~\eqref{eq:tilde-projector} are identical, and Eq.~\eqref{eq:tilde-psi} reduces to Eq.~\eqref{eq:psi}.
Consequently, the probabilities defined in Eqs.~\eqref{eq:P-m} and~\eqref{eq:tilde-P-m-tilde-m} coincide up to the prefactor.
This establishes Eq.~\eqref{eq:P-equals-tilde-P} for the case in which $m_i = \widetilde{m}_i$ for all $i \in \mathsf{U}$.

To establish Eq.~\eqref{eq:P-equals-tilde-P} in full generality, we now consider the case in which $m_i = -\widetilde{m}_i$ for a subset of unreliable measurements $i \in \mathsf{M} \subseteq \mathsf{U}$, while $m_i = \widetilde{m}_i$ for all $i \in \mathsf{U} \setminus \mathsf{M}$.
We show that the set of measurements for which assigned and actual outcomes differ can be reduced iteratively until $\mathsf{M} = \emptyset$.
Specifically, for $i \in \mathsf{M}$, the projectors~\eqref{eq:projector} and~\eqref{eq:tilde-projector} project onto orthogonal subspaces,
\begin{equation}
    \widetilde{P}_i = \frac{1}{2} \left( 1 - m_i M_i \right) = 1 - P_i.
\end{equation}
As we now demonstrate, the projectors $\widetilde{P}_i$ appearing in Eq.~\eqref{eq:tilde-psi} can be replaced by $P_i$ without affecting the distribution of assigned measurement outcomes.
This is accomplished by iteratively ``flipping'' the measurements $M_i \to -M_i$ for each $i \in \mathsf{M}$.

Flipping measurements is accomplished by exploiting stabilizers of the initial state $\ket{\psi_0}$.
Recall that each logical computational qubit is initially prepared in either the $\ket{+}$ or $\ket{0}$ state.
Consequently, the stabilizers of the initial state are products of Pauli $X$ and $Z$ operators.
As shown in Ref.~\cite{CainFastCorrelatedDecoding2025}, it is possible to choose a basis of stabilizers such that, for every unreliable measurement $M_i$ with $i \in \mathsf{U}$, there exists a unique stabilizer $G_i$ whose forward propagation through the adaptive circuit to layer $i'$,
\begin{equation}
    G_{i, i'} = U_{i'} \dotsb U_1 G_i U_1^{\dagger} \dotsb U_{i'}^{\dagger},
\end{equation}
anticommutes with $M_i$ while commuting with all other unreliable measurements $M_{i'}$ with $i' \in \mathsf{U} \setminus \{ i \}$,
\begin{equation}
    \left\{ G_{i, i}, M_i \right\} = \left[ G_{i, i'}, M_{i'} \right] = 0.
\end{equation}
In terms of projection operators, these relations can be rewritten as
\begin{equation}
    \label{eq:commutation-stabilizer-projector}
    \begin{split}
        \left( 1 - P_i \right) G_{i, i} & = G_{i, i} P_i,     \\
        P_{i'} G_{i, i'}                & = G_{i, i'} P_{i'}.
    \end{split}
\end{equation}
These commutation relations show that the stabilizer $G_i$ can be used to flip the measurement $M_i$ appearing in the projector $\widetilde{P}_i$, while leaving all other projectors $\widetilde{P}_{i'}$ with $i' \in \mathsf{M} \setminus \{ i \}$ unchanged.
To see that the projectors $\widetilde{P}_{i'}$ with $i' \in \mathsf{R}$ are likewise unaffected by commutation with $G_{i, i'}$, note first that $G_{i, i'}$ commutes with all reliable measurement products by construction.
That is, for every $i' \in \mathsf{R}$,
\begin{equation}
    \label{eq:commutation-stabilizer-reliable}
    \left[ G_{i, i'}, R_{i'} \right] = 0.
\end{equation}
Using Eq.~\eqref{eq:reliable-product-projector}, this relation can be cast into the same form as the second line of Eq.~\eqref{eq:commutation-stabilizer-projector}.

Now consider $i \in \mathsf{M}$ and use $G_i \ket{\psi_0} = \ket{\psi_0}$ in Eq.~\eqref{eq:tilde-psi}.
We commute $G_i$ to the left using $U_{i'} G_{i, i' - 1} = G_{i, i'} U_{i'}$.
To commute $G_{i, i'}$ through the projection operators, we use the relations in Eq.~\eqref{eq:commutation-stabilizer-projector}.
In this way, $\widetilde{P}_i = 1 - P_i$ is transformed into $P_i$, while all other projectors with $i' \neq i$ remain unchanged.
We then commute the stabilizer all the way to the left.
Since the stabilizer is unitary, this does not affect the probability distribution~\eqref{eq:tilde-P-m-tilde-m}.
Thus, $\mathsf{M}$ is effectively replaced by $\mathsf{M} \setminus \{ i \}$.
Repeating this procedure for all remaining elements of $\mathsf{M} \setminus \{ i \}$ eventually yields $\mathsf{M} = \emptyset$.

Finally, we iteratively use $m_{i'} P_{i'} = M_{i'} P_{i'}$ to bring the projectors~\eqref{eq:reliable-product-projector} with $i \in \mathsf{R}$ into the standard form of Eq.~\eqref{eq:projector}.
This establishes Eq.~\eqref{eq:P-equals-tilde-P}.
Hence, the outcomes of unreliable measurements can be assigned uniformly at random, provided that the assigned values of reliable products are chosen to coincide with their actual values.

\subsection{Example: Bell state}

We illustrate the equivalence of sampling assigned and actual measurement outcomes with two concrete examples.
The first one is the preparation of a Bell state and subsequent measurement of correlations in the $Z$ basis as shown in \cref{fig:bell-logical}.
In the notation used above, the initial state is $\ket{\psi_0} = \ket{+} \ket{0}$.
The circuit has $N_M = 2$ layers with unitary gates and measurements given by
\begin{equation}
    U_1 = {\cx}_{1,2}, \quad M_1 = Z_1, \quad U_2 = 1, \quad M_2 = Z_2.
\end{equation}
Applying the CNOT gate to the initial state yields the Bell state $\ket{\Phi} = U_1 \ket{\psi_0}$.

The backpropagations of $M_1$ and $M_2$ terminate as $Z_1$ on $\ket{+}$ and are thus unreliable.
However, $R_2 = M_1 M_2$ is a reliable measurement product.
The stabilizer that flips $M_1 = Z_1$ is given by $G_1 = X_1$, with forward propagation $G_{1, 1} = U_1 G_1 U_1^{\dagger} = X_1 X_2$ and $G_{1, 2} = G_{1, 1}$.
The forward propagation commutes with the reliable measurement product, $\left[ G_{1, 2}, R_2 \right] = 0$.

Let us first consider sampling the outcomes of $M_1$ and $M_2$ directly.
The joint distribution of measurement outcomes reads
\begin{equation}
    p(\mathbf{m}) = \norm{P_2 P_1 \ket{\Phi}}^2 = \frac{1}{2} \delta_{m_1, m_2}.
\end{equation}
Instead, distinguishing between assigned and actual measurement outcomes, the joint distribution takes the form
\begin{equation}
    \widetilde{p}(\mathbf{m}, \widetilde{\mathbf{m}}) = \frac{1}{2} \norm{\widetilde{P}_2 \widetilde{P}_1 \ket{\Phi}}^2 = \frac{1}{4} \delta_{\widetilde{m}_1, \widetilde{m}_2},
\end{equation}
where the assigned outcome of the second measurement is $m_2 = m_1 \widetilde{m}_1 \widetilde{m}_2$ as follows from the condition that the reliable product $R_2 = M_1 M_2$ should take the same value when expressed in terms of assigned or actual measurement outcomes.

To check that the distributions for the two strategies agree, we first consider the case $m_1 = \widetilde{m}_1$.
Then also $m_2 = \widetilde{m}_2$, and we directly obtain $p(\mathbf{m}) = 2 \widetilde{p}(\mathbf{m}, \mathbf{m})$.

Let us now consider a mismatch between the assigned and actual outcomes of the first measurement, $m_1 = - \widetilde{m}_1$.
Expressing the projector $\widetilde{P}_2$ in terms of the reliable product $R_2$ as in Eq.~\eqref{eq:reliable-product-projector}, we write the probability distribution as
\begin{multline}
    \widetilde{p}(\mathbf{m}, \widetilde{\mathbf{m}}) = \frac{1}{32} \norm{\left( 1 + m_1 m_2 R_2 \right) \left( 1 - m_1 M_1 \right) \ket{\Phi}}^2.
\end{multline}
The minus sign in the projector containing $M_1$ can be flipped by inserting the stabilizer of the Bell state, $\ket{\Phi} = G_{1, 1} \ket{\Phi}$, and using that $M_1 G_{1, 1} = - G_{1, 1} M_1$ and $M_2 G_{1, 2} = G_{1, 2} M_2$.
Finally, in the product with the projector $P_1 = \frac{1}{2} \left( 1 + m_1 M_1 \right)$, the reliable product can be rewritten as $R_2 P_1 = m_1 M_2 P_1$, which again establishes the equality between the two sampling strategies.

In this example, the first measurement is unreliable with equiprobable outcomes.
Accordingly, we can assign an outcome independently from the actual outcome.
Also the second measurement is unreliable with equiprobable outcomes according to the marginal distribution $p(m_2) = \sum_{m_1 = \pm 1} p(m_1, m_2) = 1/2$.
However, during execution of the circuit, the outcome of the second measurement is conditioned on the outcome of the first measurement.
Bell-state correlations result in the second measurement being deterministic: $p(m_2 \vert m_1) = p(m_1, m_2)/p(m_1) = \delta_{m_1, m_2}$.
When sampling assigned measurement outcomes, these correlations are accounted for by requiring the assigned and actual values of the reliable product $R_2$ to agree, $m_1 m_2 = \widetilde{m}_1 \widetilde{m}_2$.

\subsection{Example: magic state injection}

Our second example is the application of a $T$ gate on a qubit in the state $\ket{+}$ through magic state injection,
\begin{equation}
    \begin{quantikz}
        \lstick{$\ket{+}$} & \ctrl{1} & \gate{S} & \meter{M_2 = X_1} \\
        \lstick{$\ket{T}$} & \targ{}  & \gate[priority label=below, ps=meter, disable auto height]{M_1 = Z_2} \wire[u][1]{c}
    \end{quantikz}.
\end{equation}
Here, the initial state is $\ket{\psi_0} = \ket{+} \ket{T}$.
The first layer of the circuit consists of a CNOT gate and a measurement of the auxiliary qubit in the $Z$ basis,
\begin{equation}
    U_1 = {\cx}_{1,2}, \qquad M_1 = Z_2.
\end{equation}
The measurement $M_1$ is unreliable.
In the second layer of the circuit, feedforward is applied based on the assigned outcome $m_1$, and the computational qubit is measured in the $X$ basis,
\begin{equation}
    U_2(m_1) =
    \begin{cases}
        1   & m_1 = + 1, \\
        S_1 & m_1 = - 1,
    \end{cases}
    \qquad
    M_2 = X_1.
\end{equation}
The second measurement is reliable for $m_1 = + 1$ and unreliable for $m_1 = - 1$.
In the latter case, however, the product $R_2 = M_1 M_2 = X_1 Z_2$ is reliable.
The stabilizer that flips $M_1 = Z_2$ is $G_1 = X_1$ with forward propagation $G_{1, 1} = U_1 G_1 U_1^{\dagger} = X_1 X_2$ and
\begin{equation}
    G_{1, 2} =
    \begin{cases}
        X_1 X_2   & m_1 = + 1, \\
        - Y_1 X_2 & m_1 = - 1.
    \end{cases}
\end{equation}
For $m_1 = + 1$, $G_{1, 2}$ commutes with the reliable measurement $M_2 = X_1$; for $m_1 = - 1$, $G_{1, 2}$ commutes with the reliable measurement product $R_2 = X_1 Z_2$.
As in the previous example, inserting the stabilizer $G_1$ of the initial state allows us to correct a potential mismatch between the assigned and actual outcomes of the first, unreliable measurement, and thus establish the equivalence between the different sampling strategies.

Finally, it is instructive to consider the case of a computational qubit in the state $\ket{0}$, so that the first measurement $M_1 = Z_2$ is reliable.
Since the backpropagation of $M_1$ terminates as $Z_2$ on the auxiliary qubit in the state $\ket{T}$, both outcomes of the reliable measurement $M_1$ are equiprobable.
However, we are not allowed to randomly assign the outcome of $M_1$.
The reason for this is that while $\ket{T}$ states are $+ 1$ eigenstates of $T X T^{\dagger} = (X + Y)/\sqrt{2}$, this operator is not contained in the Pauli group.
Hence it cannot be used to correct a mismatch between assigned and actual outcomes of the auxiliary qubit measurement.

\begin{figure}[!h]
    \centering
    \includegraphics[width=\linewidth]{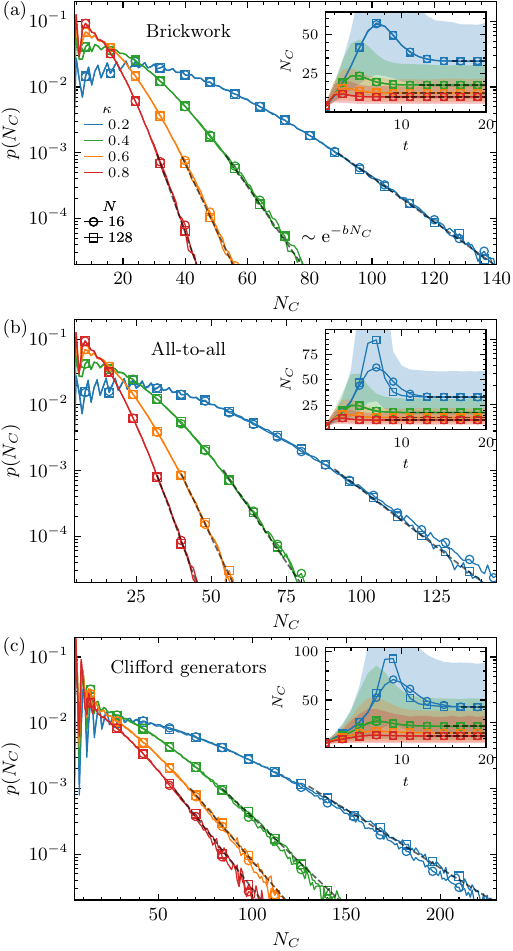}
    \phantomsubfloat{\label{fig:circuit-layouts-brickwork}}
    \phantomsubfloat{\label{fig:circuit-layouts-all}}
    \phantomsubfloat{\label{fig:circuit-layouts-clifford}}
    \caption{
        Decoding complexity for different circuit architectures.
        Main panels show the stationary distribution $p(N_C)$ of the normalized check count, while the insets show the average $N_C$ as a function of circuit layer $t$.
        Colors indicate the magic state injection probability $\kappa$, while marker shapes distinguish $N=16$ and $128$.
        (a)~Random two-qubit Clifford gates in a one-dimensional nearest-neighbor brickwork geometry; the data are the same as in \cref{fig:complexity} and are reproduced here for comparison.
        (b)~The same random two-qubit Clifford gates applied to random pairings of qubits with all-to-all connectivity.
        (c)~All-to-all circuit constructed from the generating set $\{ H, S, \cx \}$ of the Clifford group.
    }
    \label{fig:circuit-layouts}
\end{figure}

\section{Additional circuit layouts and gate ensembles}
\label{sec:other-layouts}

In our analysis of the decoding complexity in \cref{sec:compl-rand-circ}, we focused primarily on the random brickwork circuits shown in \cref{fig:circuit-model}, in which two-qubit Clifford gates act on neighboring computational qubits in a one-dimensional geometry.
Here, we show that the qualitative conclusions of this analysis are robust to changes in both the connectivity and the Clifford gate ensemble.

We compare three circuit ensembles in \cref{fig:circuit-layouts}.
The brickwork-circuit results shown in \cref{fig:complexity} are reproduced in \cref{fig:circuit-layouts-brickwork} for reference.
In \cref{fig:circuit-layouts-all}, we retain the same random two-qubit Clifford gates but replace the one-dimensional nearest-neighbor connectivity by all-to-all connectivity: in each Clifford layer, the computational qubits are randomly paired and a random two-qubit Clifford gate is applied to each pair.
\cref{fig:circuit-layouts-clifford} considers an all-to-all circuit constructed directly from Clifford generators.
Each Clifford layer consists of a single-qubit sublayer, in which one of $\{I,H,S\}$ is applied independently to every computational qubit with equal probability, and a two-qubit sublayer, consisting of CNOT gates on a random pairing of the computational qubits.
The nontrivial Clifford gates are therefore drawn from the generating set $\{H,S,\cx\}$, which admits fold-transversal implementations with surface-code logical qubits~\cite{MoussaTransversalCliffordGates2016, BreuckmannFoldTransversalCliffordGates2024}.

For all three ensembles, we consider $N=16$ and $128$ computational logical qubits and four values of the magic state injection probability $\kappa$.
The data are averaged over $N_{\mathrm{real}}=10^4$ and $10^3$ circuit realizations for $N=16$ and $128$, respectively, using the same procedure as in \cref{sec:decoding-complexity}.
Reliable measurement products are selected using the Free prescription introduced in \cref{sec:choice-product}.

Across all three circuit ensembles, we observe the same qualitative behavior found for the brickwork circuits in \cref{sec:decoding-complexity}.
After an initial transient and overshoot, the average normalized check count approaches a stationary value $N_{C, \infty}$.
At fixed $\kappa$, this stationary value becomes independent of the number of computational qubits $N$, although its value depends on the circuit ensemble through the number of checks per circuit layer as discussed below.
Likewise, the stationary distributions shown in the main panels of \cref{fig:circuit-layouts} are insensitive to $N$ and exhibit approximately exponential suppression at large $N_C$, with increasingly steep tails for increasing $\kappa$.

The stationary values $N_{C, \infty}$ can be understood using the same counting argument as in \cref{sec:decoding-complexity}.
For the brickwork and all-to-all circuits in \cref{fig:circuit-layouts-brickwork,fig:circuit-layouts-all}, the gate structure and syndrome-extraction schedule are identical and the circuits differ only in their connectivity.
The counting argument therefore applies unchanged, yielding the stationary value in \cref{eq:N-C-infty}.

\begin{figure*}[!tb]
  \centering
  \includegraphics[width=\textwidth]{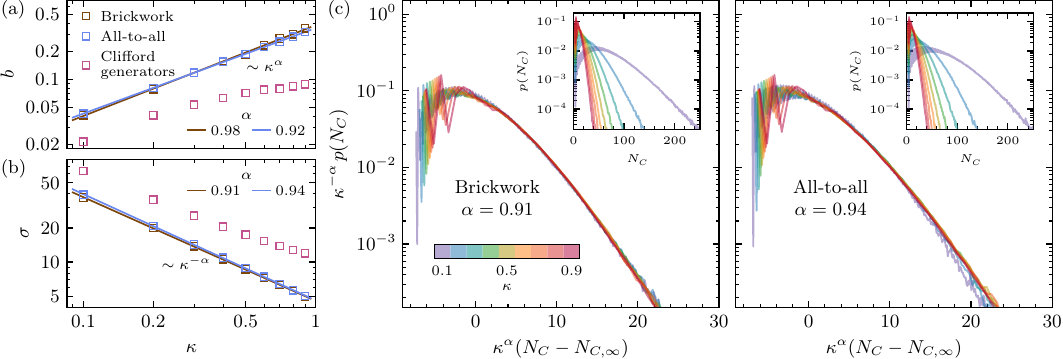}  
  \phantomsubfloat{\label{fig:scaling-b}}
  \phantomsubfloat{\label{fig:scaling-sigma}}
  \phantomsubfloat{\label{fig:scaling-distribution}}
  \vspace*{-1em}
  \caption{
    Scaling analysis of the stationary check-count distributions in \cref{fig:circuit-layouts}.
    (a)~Decay rate $b$ of the exponential tail as a function of $\kappa$.
    Power-law fits $b\sim\kappa^{\alpha}$ yield $\alpha=0.98$ for brickwork circuits and $\alpha=0.92$ for all-to-all circuits.
    (b)~Standard deviation $\sigma$ of the stationary distributions.
    Fits to $\sigma\sim\kappa^{-\alpha}$ yield $\alpha=0.91$ and $\alpha=0.94$ for brickwork and all-to-all circuits, respectively.
    Neither observable exhibits a comparable power-law dependence for Clifford-generator circuits.
    (c)~Collapse of the stationary distributions for brickwork and all-to-all circuits according to \cref{eq:scaling-form}, using the exponents obtained from the fits to $\sigma \sim \kappa^{\alpha}$ in (b).    
    Insets show the distributions before rescaling.
  }
  \label{fig:scaling}
\end{figure*}

For the Clifford-generator circuit in \cref{fig:circuit-layouts-clifford}, the Clifford part of each circuit layer instead contains separate single- and two-qubit sublayers, contributing $2N$ checks each.
Following the same counting argument as in \cref{sec:decoding-complexity}, the stationary value is therefore modified to
\begin{equation}
    N_{C,\infty}
    =
    \frac{8+3\kappa}{\kappa},
\end{equation}
which accounts for the larger stationary check counts observed in \cref{fig:circuit-layouts-clifford}.

For the all-to-all circuits in \cref{fig:circuit-layouts-all,fig:circuit-layouts-clifford}, a weak dependence on $N$ remains visible during the transient overshoot at the smallest injection probability, $\kappa=0.2$.
Instead, for larger $\kappa$, the results for $N=16$ and $128$ agree closely throughout the evolution.
Therefore, we expect also the residual transient dependence at $\kappa=0.2$ to disappear with increasing $N$.

In summary, these results support the conclusion that the bounded decoding complexity found in \cref{sec:decoding-complexity} is not specific to the one-dimensional brickwork geometry or to the particular Clifford gate ensemble used there.

\section{Scaling form of the stationary distribution}
\label{sec:scal-form-stat}

The one-parameter scaling form~\eqref{eq:scaling-form} implies that the decay rate $b$ of the exponential tail and the standard deviation $\sigma$ of the stationary distribution scale as $b \sim \kappa^{\alpha}$ and $\sigma \sim \kappa^{-\alpha}$, respectively.
Here, we test these predictions for the circuit ensembles introduced in \cref{sec:other-layouts} and examine the resulting collapse of the full distributions.

\Cref{fig:scaling-b} shows that the tail decay rates exhibit approximate power-law behavior for both brickwork and all-to-all circuits, which are composed of random two-qubit Clifford gates and differ only in their connectivity.
The fitted exponents are $\alpha = 0.98$ and $\alpha = 0.92$, respectively. By contrast, no comparable power-law dependence is observed for Clifford-generator circuits.

Extracting the decay rate $b$ requires fitting the sparsely sampled large-$N_C$ tails, making the results sensitive to statistical noise.
We therefore also consider the standard deviation $\sigma$, which provides a more robust measure of the width of the distribution.
As shown in \cref{fig:scaling-sigma}, we find approximate power-law behavior $\sigma \sim \kappa^{-\alpha}$ for brickwork and all-to-all circuits, with exponents $\alpha = 0.91$ and $\alpha = 0.94$, respectively.
Taken together, the fitted exponents favor a nontrivial value $\alpha < 1$, with estimates in the range $0.9 \lesssim \alpha \lesssim 1$.
However, their spread is too large to resolve a possible dependence on circuit connectivity or to conclusively exclude $\alpha = 1$.

Finally, using the exponents obtained from the standard deviations, we rescale the stationary distributions according to \cref{eq:scaling-form}.
As shown in \cref{fig:scaling-distribution}, the data collapse well for both brickwork and all-to-all circuits over a broad range of the scaling variable.
The remaining deviations are concentrated at small values of the scaling variable, where details of the microscopic circuit structure may become important.

\newpage
\bibliography{library}

\end{document}